\documentclass[twocolumn]{aa}
\usepackage{natbib}

\usepackage{txfonts}
\usepackage{natbib}

\usepackage[FIGTOPCAP]{subfigure}
\usepackage{threeparttable}
\usepackage{tabularx}
\usepackage{booktabs}
\usepackage{longtable}
\usepackage{color}
\usepackage[dvipsnames]{xcolor}

\usepackage{hyperref}
\hypersetup{
    colorlinks   = true,
    citecolor=blue,
}

\def\srv{$\sigma_\mathrm{RV}$}

\def\halp{H$\alpha$}

\def\hbet{H$\beta$}

\def\hea{He\,{\sc i}}

\def\oi{O\,{\sc i}}

\def\fei{Fe\,{\sc i}}
\def\feii{Fe\,{\sc ii}}

\def\srv{$\sigma_\mathrm{1D}$}

\def\ngc6357{NGC\,6357}
\def\g333{G333.6-0.2}

\defcitealias{Ramirez-Tannus2017}{RT17}
\newcommand{\RT}{\citetalias{Ramirez-Tannus2017}}

\begin{document}

   \title{Spectroscopic variability of massive pre-main-sequence stars in M17\thanks{Based on observations collected at the European Southern Observatory at Paranal, Chile (ESO program 0103.D-0099, 60.A-9404(A), 085.D-0741,
089.C-0874(A), and 091.C-0934(B).}}

   \author{A.R. Derkink
          \inst{1}
    \and
    M.C. Ram\'{i}rez-Tannus
    \inst{2}
  	\and
   	L. Kaper
   	\inst{1}
    \and
       	A. de Koter
   	\inst{1,3}
    \and
   	F. Backs 
   	\inst{1}
        \and
   	J. Poorta
    \inst{1}
   	\and
    M.L. van Gelder
    \inst{4}}

   \institute{Anton Pannekoek Institute, University of Amsterdam,
              Sciencepark, A-1180 Amsterdam\\
              \email{a.r.derkink@uva.nl}
              \and
              Max Planck Institute for Astronomy, K\"{o}nigstuhl 17, 69117, Heidelberg, Germany
              \and
   	Instituut voor Sterrenkunde, KU Leuven, Celestijnenlaan 200D bus 2401,
   	3001 Leuven, Belgium
    \and 
     Leiden Observatory, Leiden University, PO Box 9513, 2300 RA Leiden, The Netherlands}

  \date{}

 \abstract
   {It is a challenge to study the formation process of massive stars: their formation time is short, they are few, often deeply embedded, and at relatively large distances. Our strategy is to study the outcome of the star formation process and to look for signatures remnant of the formation. We have access to a unique sample of (massive) pre-main-sequence (PMS) stars in the giant H~{\sc ii} region M17. These PMS stars can be placed on pre-main-sequence tracks in the Hertzsprung-Russell diagram (HRD) as we can detect their photospheric spectrum, and exhibit spectral features indicative of the presence of a circumstellar disk. These stars are most likely in the final stage of formation.}
   {The aim is to use spectroscopic variability as a diagnostic tool to learn about the physical nature of these massive PMS stars. More specifically, we want to determine the variability properties of the hot gaseous disks to understand the physical origin of the emission lines and identify dominant physical processes in these disks, and to find out about the presence of an accretion flow and/or jet.}
   {We have obtained multiple-epoch (4-5 epochs) VLT/X-shooter spectra of six young stars in M17 covering about a decade; four of them are intermediate to massive PMS stars with gaseous disks. Using stacked spectra we update the spectral classification and search for the presence of circumstellar features. With the temporal variance method (TVS) we determine the extent and amplitude of the spectral line variations in velocity space. The double-peaked emission lines in the PMS stars with gaseous disks are used to determine peak-to-peak velocities, V/R-ratios and the radial velocity of the systems. Simultaneous photometric variations are studied using VLT acquisition images.}
  {From detailed line identification in the PMS stars with gaseous disks, we identify many (double-peaked) disk features, under which a new detection of CO bandhead and C\,{\sc i} emission. In three of these stars we detect significant spectral variability, mainly in lines originating in the circumstellar disk, in a velocity range up to $320$~km~s$^{-1}$, which exceeds the rotational velocity of the central sources. The shortest variability timescale is of the order of a day; also long-term (months, years) variability is detected. In two PMS stars the ratio between the blue and red peaks shows a correlation with the peak-to-peak velocity, possibly explained by a spiral-arm structure in the disk.}
  {The PMS stars with variability are at similar positions in the HRD but show significant differences in disk lines and variability. The extent and timescale of the variability differs for each star and per line (sets), showing the complexity of the regions where the lines are formed. We find indications for an accretion flow, slow disk winds and/or disk structures in the hot gaseous inner disks and do not find evidence for close companions or strong accretion bursts as the cause of the variability in these PMS stars.}


   \maketitle
%

\section{Introduction}\label{sec:intro}
Since the advent of 8–10m telescopes our understanding of the formation process of massive stars has significantly improved. Massive stars likely form through disk accretion \citep[e.g. ][]{kuiper2018}, similar to low-mass stars. Near-infrared (NIR) and optical spectra of massive young stellar objects (MYSOs) include characteristic features (CO bandhead emission, double-peaked emission lines, NIR excess), demonstrating the presence of a circumstellar disk \citep{Hanson1997, bik2006}. Some MYSOs also reveal photospheric absorption lines confirming their PMS nature \citep{Ochsendorf2018} consistent with model predictions \citep[e.g., ][]{hosokawa2011}. Others show jet emission lines \citep{ellerbroek2011} indicative of active accretion, but expose no photospheric signatures. The detailed spectroscopic properties \citep[e.g., the presence of CO bandhead emission; ][]{ilee2018, poorta2023} vary from object to object, indicating substantial differences in disk properties and/or PMS phase.

Variability is a key diagnostic in studying astrophysical phenomena. Variability studies have made an important contribution to understanding the star formation process and the PMS evolution of young low-mass stars, by identifying the key physical processes \citep[e.g.,][]{cabrit1990, muzerolle2004, audard2014, contreras-pena2017, Fischer2022}. The optical/infrared variability in low-mass YSOs reveals phenomena such as accretion events in the disk, magnetospheric activity, spots, and flares \citep[e.g.,][]{carpenter2001, morales-calder2011}. In intermediate mass stars, e.g., the Herbig Ae/Be stars, variability studies provide information on the physical structure of the disk \citep{ilee2014}, the presence of moving structure within jets \citep{Ellerbroek2014}, the accretion rate \citep[e.g.,][]{mendigutia2020, Moura2020}, and the occurrence of close binaries \citep{baines2006}. 

It is difficult to study the PMS phase of massive stars due to the observational challenge to detect them in this early phase: massive stars are rare, the formation timescales are short, i.e. $10^{4}$ yr rather than $10^{7}$ yr for a solar-type star, they are deeply embedded in their parental cloud (with $A_{\rm V} \sim 10 - 100$~mag) and are located at relatively large distances. Therefore, little is known about the temporal behavior of MYSOs. \cite{kumar2016} identified high-amplitude infrared variable stars in a NIR photomeric survey, from which 13 sources qualify as MYSO. They conclude that the large variations are due to episodic accretion events. \cite{caratti2017} and \cite{uchiyama2019} study the temporal behavior of an early class MYSO and reveal high-mass accretion events, which are investigated with the use of numerical models of \cite{meyer2017}. \cite{Mendigutia2011} and \cite{Moura2020} study the temporal behavior in intermediate mass YSOs, likely caused by mass accretion events and disk winds. 

So far, optical and near-infrared variability studies of MYSOs have mainly been performed on photometric data. But we are now in the position to search for optical and NIR spectroscopic variability in a sample of intermediate to high mass PMS stars characterized by \cite{Ramirez-Tannus2017} (after this: \citetalias{Ramirez-Tannus2017}). These PMS stars are embedded in the young ($\sim1$~Myr) massive star forming region M17, the nearest giant H\,{\sc ii} region at $1.68_{-0.11}^{0.13}$\,kpc \citep{Kuhn19}.
Interestingly, they are displaying a photospheric spectrum at optical and NIR wavelengths while still surrounded by a disk. Through quantitative spectroscopy, they were placed on PMS evolutionary tracks allowing to constrain their age. Also other properties such as the mass (estimated based on the position in the HRD), the projected rotational velocity and extinction parameters
have been determined.

Based on single-epoch spectra \RT\ concluded that the detected disks are likely remnant of the formation process. It may well be that these remnant disks are currently being destroyed by the increasing ultraviolet (UV) flux of the contracting star and/or the onset of the radiation-driven wind \citep{hollenbach1994, hollenback2000, bik2006}. A spectroscopic variability study could reveal information on these dynamic processes affecting the structure of the disk. 

The aim of this paper is to search for variations in multi-epoch spectra of intermediate to massive PMS stars, and, if detected, to use this variability as a diagnostic tool to better understand the dominant physical processes involved in this phase of the formation of massive stars. This information could be used to map the disk morphology, to identify changes in the disk structure, and to study the accretion process \citep{Mendigutia2011, Mendigutia2013, pogodin2022}. In this paper, we study six stars in M17, which are identified as intermediate to high mass PMS stars by \RT. In four of these stars which have with gaseous disks, we detect and analyze spectroscopic variations. We characterize the incident rate and extent of the variability, as well as the timescale(s) associated with this behavior.

In the next section we introduce the sample of PMS stars, describe the data reduction procedure and present the photometric variability in three targets. In Sec.~3 we update the spectral classification and in Sec.~4 we introduce the methods to study spectral lines formed in the circumstellar material and spectroscopic variability in the stars. In Sec.~5 and 6 we identify circumstellar lines and double-peaked line properties, and in Sec.~7 we present the spectroscopic variability study, applying a time variance analysis to  quantify the extent and amplitude of the variability. In the last sections we summarize our conclusions and discuss the results in a broader perspective.

\begin{figure}
\centering
\includegraphics[width=\hsize]{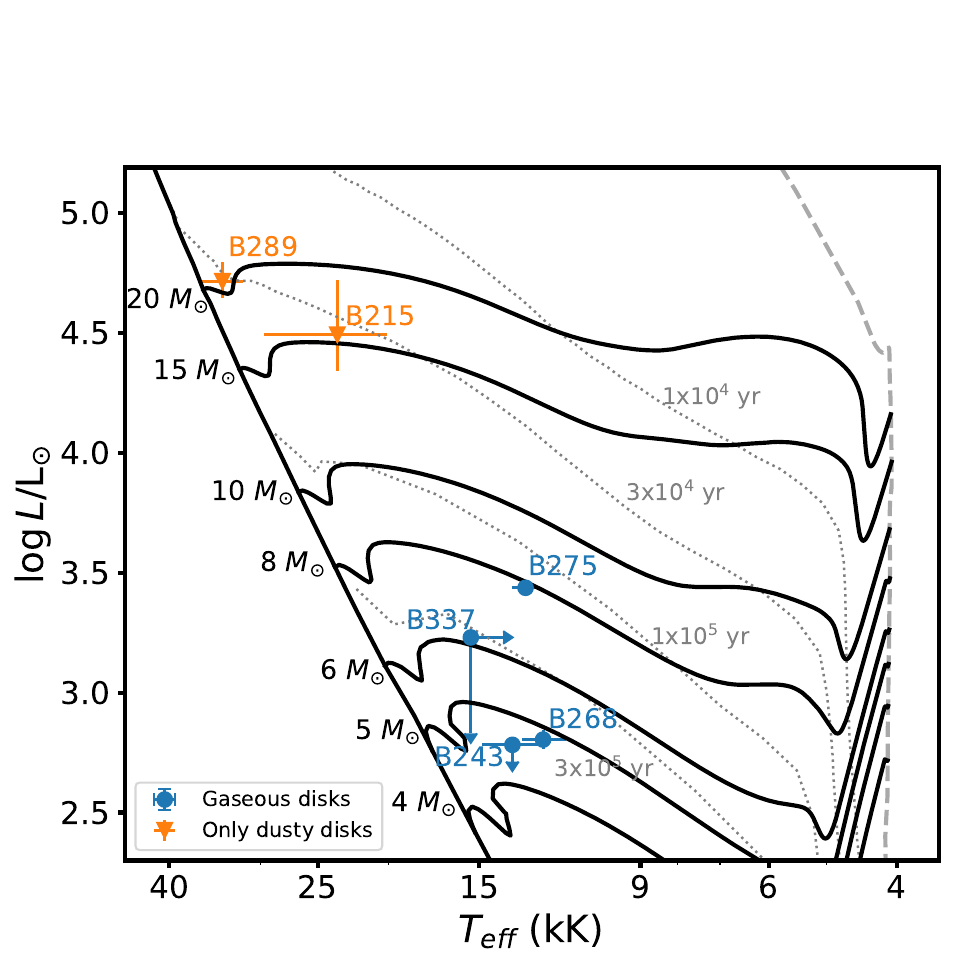}
  \caption{Hertzsprung-Russell diagram of the stars in M17 with PMS evolutionary tracks from MIST \citep[solid black lines;][]{dotter2016,choi2016,paxton2011,paxton2013,paxton2015}. The blue dots represent the PMS stars with IR excess and emission lines in their X-Shooter spectrum. The orange triangles represent the PMS stars with no emission lines in the X-Shooter spectrum but with IR excess $\lambda$ > 2.3\,$\mu$m \citep{Chini2005}. The gray striped line is the birth line and the straight black line to the left of the figure is the zero age main sequence (ZAMS). The dotted gray lines show the isochrones with ages indicated in a similar color in the figure.
          }
     \label{HRD}
\end{figure}

\renewcommand{\arraystretch}{1.8}
\begin{table*}[]
\caption{The updated spectral type and stellar parameters of our sample stars.}
\centering
\begin{tabular}{@{}lllrlrrrll@{}}
\toprule
\textbf{} & Sp. Type  &  Sp. Type & T$_{\mathrm{eff}}^{\rm a}$ & log\,g $^{\rm a}$ & $v$\,sin$i$ $^{\rm a}$ & R & $A_\mathrm{V}$ $^{\rm b}$ & log L/L$_{\odot}$  & distance $^{\rm c}$ \\
    &  This work & \RT  & K  & cm\,s$^{-2}$ & km\,s$^{-1}$ & R$_{\odot}$   & mag  &          & pc    \\ \midrule \midrule

B215 & B1 V &      B0-B1 V &            $23500^{+6100}_{-3450}$  & $3.67^{+0.5}_{\downarrow}$    & $208^{+40}_{-38}$     & $10.6^{+7.5}_{-3.2}$   & $9.1^{\pm0.3}$     & $4.49^{+0.4}_{-0.5}$ &   $1687^{+227}_{-178}$ \\
\textbf{B243} & B9 III  & B8 V   &      $13500^{+1350}_{-1250}$   & $4.34^{\uparrow}_{-0.8}$    & $110^{+106}_{\downarrow}$     & $4.5^{+0.6}_{-0.6}$   & $7.9^{\pm0.1}$      & $2.78^{+0.13}_{-0.14}$ &   1675\\
\textbf{B268} & B9 III  &  B9-A0   &    $12250^{+850}_{-1000}$   & $3.99^{\uparrow}_{-0.8}$    & $36^{+126}_{\downarrow}$     & $5.6^{+0.7}_{-0.6}$   & $7.5^{\pm0.1}$       & $2.80^{+0.10}_{-0.12}$ &   1675\\
\textbf{B275} & B7 III  & B7 III   &    $12950^{+550}_{-650}$   & $3.39^{+0.06}_{-0.11}$    & $88^{+68}_{\downarrow}$      & $10.4^{+1.0}_{-0.9}$  & $7.4^{\pm0.1}$       & $3.44^{+0.07}_{-0.10}$ &   1675 \\
B289 & O9.5 V  &  O9.7 V   &            $33800^{+2500}_{-2150}$   & $3.93^{+0.35}_{\downarrow}$    & $154^{+38}_{-34}$     & $6.7^{+0.4}_{-1.1}$   & $8.13^{\pm0.03}$      & $4.72^{+0.10}_{-0.09}$ &  $1694^{+105}_{-76}$\\
\textbf{B337} & late-B  &  late-B  &    13000$^{+3000}_{-3000}$  & -       & -       & 7.3   & $15.5$      & 3.23 &   1675\\ \bottomrule
\end{tabular}
\begin{tablenotes}
    \small
    \item a. Values are adopted from \RT\ b. \cite{backs:inpress-a} c. When available, the distance is taken from \cite{Gaia2022} otherwise we use the average distance to M17 of $1675^{+19}_{-18}$\,pc from \cite{stoop_inprep}
\end{tablenotes}

\label{tab:MYSO_spec}
\end{table*}

\renewcommand{\arraystretch}{1.}

\section{Observations, data reduction and sample selection}
\subsection{Sample selection}\label{sec:sample}
Six stars, B215, B243, B268, B275, B289 and B337 are selected from the spectroscopic monitoring campaign of OB stars in M17 with the X-shooter spectograph on the ESO Very Large Telescope (VLT) (Run 0103.D-0099, P.I. Ramírez-Tannus). All targets have three to five archival X-shooter spectra (see Tab.~\ref{tab:runs}), used by \RT\ to investigate the pre-main-sequence nature based on their disk signatures and/or infrared excess. 

Four of the targets (B243, B268, B275 and B337) exhibit a gaseous circumstellar disk based on detected double-peaked emission lines in the spectrum, CO bandhead emission (except for B337) and IR excess originating from disks with gas and dust. The gas in the disk of three of these four systems has been modeled and observed at relatively high inclination of $\sim$\,70$^{\circ}$ for B243 \citep{Backs2023}; $\sim$\,80$^{\circ}$ for B268, and $\sim$\,60$^{\circ}$ for B275 \citep{poorta2023}, i.e., we apparently observe the disks fairly edge on.

The two other targets (B215 and B289) only show IR excess at $\lambda$ > 2.3\,$\mu$m \citep{Chini2005} but lack emission features in the X-Shooter spectrum. \RT\ identify them as PMS stars with dusty disks, however the lack of NIR emission does not exclude presence of colder gas further out in the disk \citep[][]{Frost2021}. 

Additionally, since all targets also show photospheric absorption lines, \RT\ placed the sample in the HRD. The effective temperatures (T$_{\mathrm{eff}}$) of the stars in Fig.~\ref{HRD} are taken from \RT\ that used a fitting algorithm to compare \ion{H}{i}, \ion{He}{i}, and \ion{He}{ii} lines with the non-LTE stellar atmosphere code FASTWIND \citep{Puls2005,riverogonzalez2012} (Table \ref{tab:MYSO_spec}). We update the luminosities of the stars in the HRD with the most recent distance and extinction parameters. The luminosity (L) is determined based on the V-band magnitudes reported in \RT, bolometric corrections for PMS stars in \cite{Pecaut2013} and distances reported by the \cite{Gaia2022}. If the sources are not listed in the Gaia DR3 catalog, we adopted the average distance to M17 $1675^{+19}_{-18}$\,pc reported by \cite{stoop_inprep} based on the Gaia DR3 parallax of 42 members of the central cluster of M17. The spectra were dereddened by \cite{backs:inpress-a} using Castelli \& Kurucz models \citep{castelli2003} and the \cite{fitzpatrick1999} extinction law, since \cite{Ramirez-Tannus2018} showed that this is a better choice for a region with relatively large R$_{\rm{v}}$ values like M17. For B337, the most embedded star whose blue part of the spectrum is obscured, L and T$_{\mathrm{eff}}$ are determined based on its spectral type. 

Different to \RT, where the authors assumed a distance of 1980\,pc \citep{Xu2011}, B243 and B268 are now closer towards the PMS track of a 5\,M$_{\mathrm{\odot}}$ star instead of a 6\,M$_{\mathrm{\odot}}$ star, B337 is close to the 6\,M$_{\mathrm{\odot}}$ track rather than the track of a 7\,M$_{\mathrm{\odot}}$ star, and B215 and B289 are both between the tracks of a 15 and 20\,M$_{\mathrm{\odot}}$ star. The PMS stars with gaseous disks are at the high mass end of the Herbig Be stars. 

\subsection{Spectroscopy}
For the six targets, optical to NIR observations (300-2500 nm) have been obtained with the X-shooter spectrograph mounted on UT2 of the VLT \citep{Vernet2011}. This study is based on 3-6 epochs in the period 2009-2019. The spectra presented in \RT\ comprise the first, and in some cases, the second epoch between 2009-2013. An additional 2-3 epochs were obtained in the summer of 2019 as part of a larger program in search for close binaries among the young OB stars in M17 (ESO program 0103.D-0099). We refer to Tab.~\ref{tab:runs} for a log of the observations and to Tab.~\ref{obslog} for a detailed overview. 

The X-shooter observations were performed under good weather conditions with seeing ranging between 0.5\arcsec\ and 1.2\arcsec\ and clear sky. The slit width used in the UVB arm (300-590 nm) is 1\arcsec (R 5100), except for the 2012 spectrum of B289 and the science verification spectrum of B275 in 2009; taken with a slit width of 0.8\arcsec\ (R 6200) and 1.6\arcsec\ (R 3300), respectively. For the VIS arm (550-1020 nm) a slit width of 0.9\arcsec\ (R 8800) is used, except for the 2012 spectrum of B289, which has a slit width of 0.7\arcsec\ (R 11,000). The NIR arm (1000-1480 nm) observations have been taken before 2019 with a slit width of 0.4\arcsec\ (R 11,300) and in 2019 with a slit width of 0.6\arcsec\ (R 8100). The exception for the NIR arm is the 2009 spectrum of B275 with a slit width of 0.9\arcsec\ (R 5600). All spectra were taken in nodding mode. 

\begin{table}[]
\caption{Overview of X-shooter epochs of the sample in M17.}
\setlength{\tabcolsep}{3pt}
\begin{tabular}{ll | llllll}
\toprule
Program   & Year & B215 & B243 & B268 & B275 & B289 & B337 
\\ \midrule \midrule
60.A-9404 & 2009 &      &      &      & x    &      &      \\
085.D-0741 & 2010     &      &      &      &      & x    &      \\
089.C-0874 & 2012    & x    & x    & xx   &      &      &      \\
091.C-0934 & 2013    &      & x    & x    &      &      & x    \\
0103.D-0099 & 2019  & xxx  & xxx  & xxx  & xxx  & xx   & xx  \\
\midrule
\textbf{Total epochs}& & 4 & 5& 6 & 4 & 3 & 3\\
\bottomrule
\end{tabular}
\begin{tablenotes}
    \small
    \item \textbf{Notes.} The 'x' represents an epoch during that year. A detailed log of observations can be found in Tab.\,\ref{obslog}. 
\end{tablenotes}
\label{tab:runs}
\end{table}

\subsubsection{Data reduction}
The spectroscopic data are reduced with the ESO X-shooter pipeline 3.3.5 \citep{Modigliani2010}. This pipeline consists of a bias subtraction, a flat field correction and a wavelength calibration. The flux calibration is done with spectrophotometric standards from the ESO database. The initial reduction was done with the xsh\textunderscore scired\textunderscore slit\textunderscore nod recipe provided in the software package from ESO. Furthermore, the Molecfit 1.5.9 tool is used to correct for the telluric sky lines \citep{Smette2015, Kausch2015}.

\begin{figure*}
   \centering
   \includegraphics[width=0.85\textwidth]{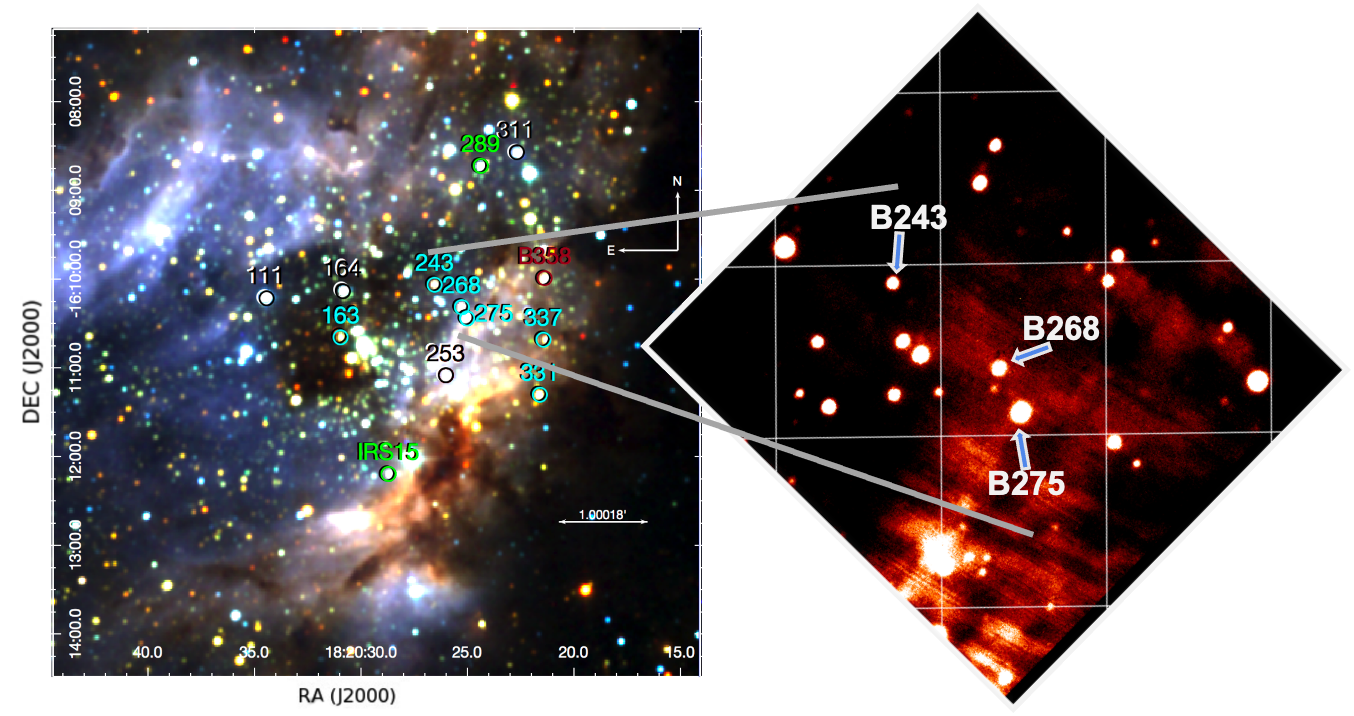}
      \caption{Location of three PMS stars with gaseous disks in M17 (figure adapted from \RT). The zoomed in region shows a VLT acquisition image taken with a 1.5'x1.5' FoV technical CCD and a z' filter. B243, B268 and B275 are in the same field of view and indicated with the arrows. Nebular emission is present most prominently near B275 and B268. 
              }
         \label{acquisition}
\end{figure*} 

\subsubsection{Nebular subtraction}\label{sec:neb}
All targets have residuals in the spectra resulting from the subtraction of strong nebular emission in the nodding mode. As the nebular line emission varies along the slit, both in strength and position, for some targets this reduction left stronger residuals overlapping the emission features, which are broader than what can be expected from the width of a nebular emission line and the spectral resolution. To mitigate these problems, a more advanced nebular fitting routine developed by \cite{gelder2020} (NebSub method) was applied to the four stars with severe nebular extinction which overlaps with line emission from the gaseous circumstellar disk. For the remaining stars the contamination does not substantially affect the analysis. In order to use this method, spectra obtained at different nodding positions are reduced individually in staring mode without sky subtraction. The two consecutive nodding observations of B268 on 06-07-2012 are added in the staring mode, leading to a total of five instead of six observations for this object. 

The NebSub method determines the strength of the forbidden nebular lines on-source and in the sky spectrum (off-source), by fitting those lines along the sky on the 2D slit, since those are assumed to solely have a nebular contribution. With help of the nebular forbidden lines on-source and off-source, the scaling in strength and position is determined. This yields the nebular model spectrum that is subtracted from the source.

The nebular lines in B268, B275 and B337 were corrected with the use of models based on double Gaussian fits to the nebular lines off source. For B243 a flat top Gaussian function was sufficient to model the nebular contamination, where only the UVB arm of X-Shooter is corrected (used for spectral classification). 
However, residuals remain in several lines with a strong nebular component. Since the nebular lines are very strong in He\,{\sc i} lines, some He\,{\sc i} with hints of circumstellar emission (e.g., the meta stable He\,{\sc i} line at 1083\,nm), still had to be excluded from the study in this paper. 

\subsection{Photometry from acquisition images}
\label{P1:sec:PhotReduction}
The VLT acquisition images, taken with a $1.5' \times 1.5'$ FoV technical CCD, provide photometric measurements in the $z^{\prime}$-band of three stars in the sample that share their field of view. The $z^{\prime}$ filter is part of the SDSS filter system and covers the wavelength band 804$-$1350\,nm. An example is shown in Fig.~\ref{acquisition}. The 17 observations of B243, B268 and B275 have the same time cadence as the spectroscopic data. The acquisition file of the science verification spectrum of B275 in 2009 is unavailable.

   \begin{figure*}
   \centering
   \includegraphics[width=0.78\hsize]{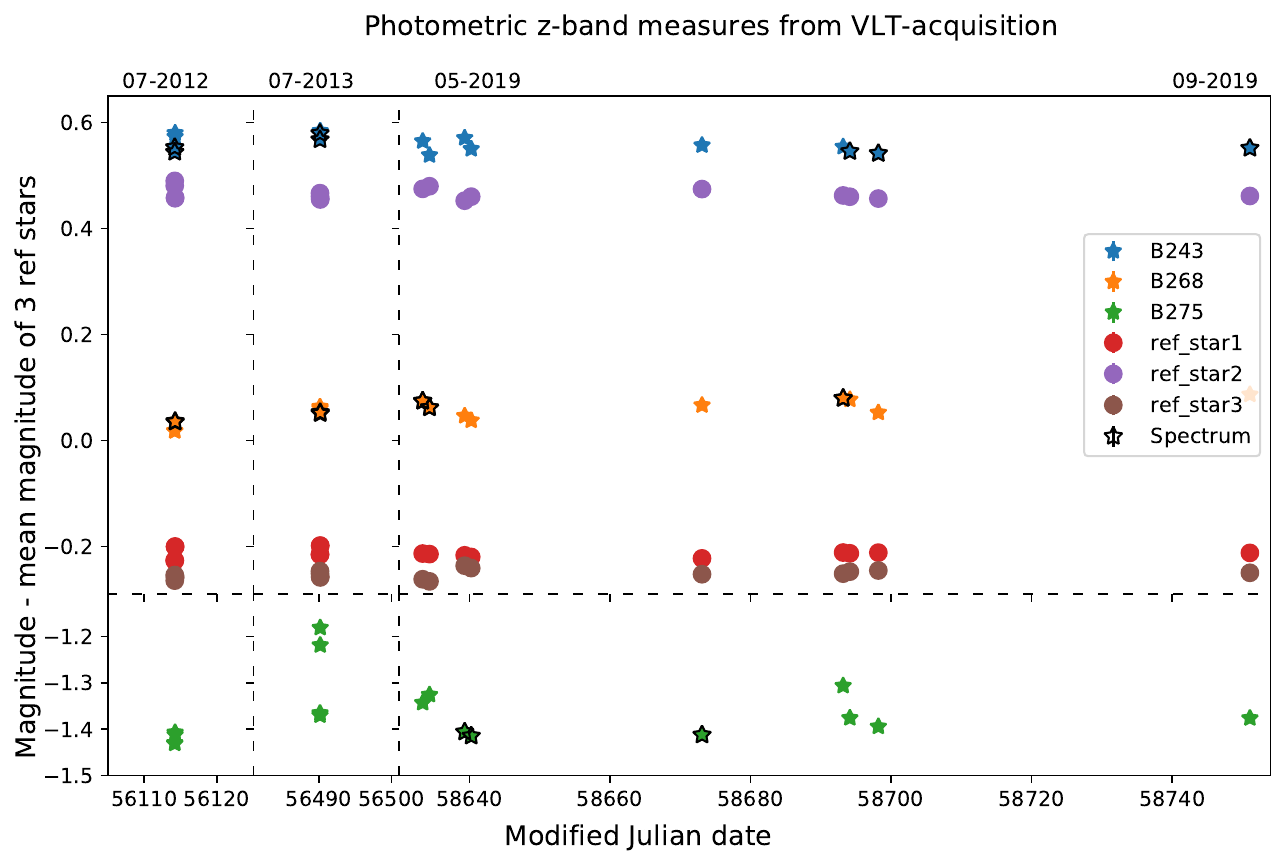}
      \caption{$z^{\prime}$-band magnitude measurements from 10\,s exposures with the VLT acquisition camera. The sample includes three reference stars (circles) from which the mean of the three has been used to correct for observational differences between the acquisitions. The black-bordered star symbols represent acquisitions of stars in the sample with an accompanying X-shooter spectrum. 
              }
         \label{photometry}
   \end{figure*} 

The acquisition images have been analyzed using the aperture photometry package Photutils \citep{larry_bradley_2020_4044744-photutils}. The magnitudes are differential magnitudes with respect to the mean of three reference stars (circles in Fig.\,\ref{photometry}) that are close to the position of the three targets, such as to compare their magnitudes over the epochs and account for possible differences due to observing conditions. 

Fig.~\ref{photometry} shows the lightcurves, where $z^{\prime}$-band magnitudes are plotted as function of Julian date. The differential $z^{\prime}$ magnitudes are 
\begin{equation}
   \Delta\,z^{\prime} = z_{\rm obs}^{\prime} - \left( \frac{\sum\,z^{\prime}_{{\rm ref},i}}{3} \right),
\end{equation}
where $z^{\prime}_{\rm obs}$ is the observed magnitude and $z^{\prime}_{{\rm ref},i}$ are the magnitudes of the reference stars $i$ (circles in Fig.~\ref{photometry}). 
The measurements for our targets are shown with a star symbol, where the black-bordered symbols denote acquisitions that have an accompanying spectrum.

For B243 and B268 the $\Delta z^{\prime}$ magnitude varies less than about $0.03$\,mag; therefore, considering potential variability of the reference stars (especially reference star 2) we conclude that no significant photometric change is detected. Only B275 reveals significant variability; $\Delta z^{\prime} = 0.2$~mag on timescales of hours in 2013 and $\Delta z^{\prime} = 0.1$~mag in 2019 on a timescale days. Unfortunately, the photometric observations showing significant variability do not have corresponding spectroscopic observations for B275.

\begin{figure*}
   \centering
   \includegraphics[width=\textwidth]{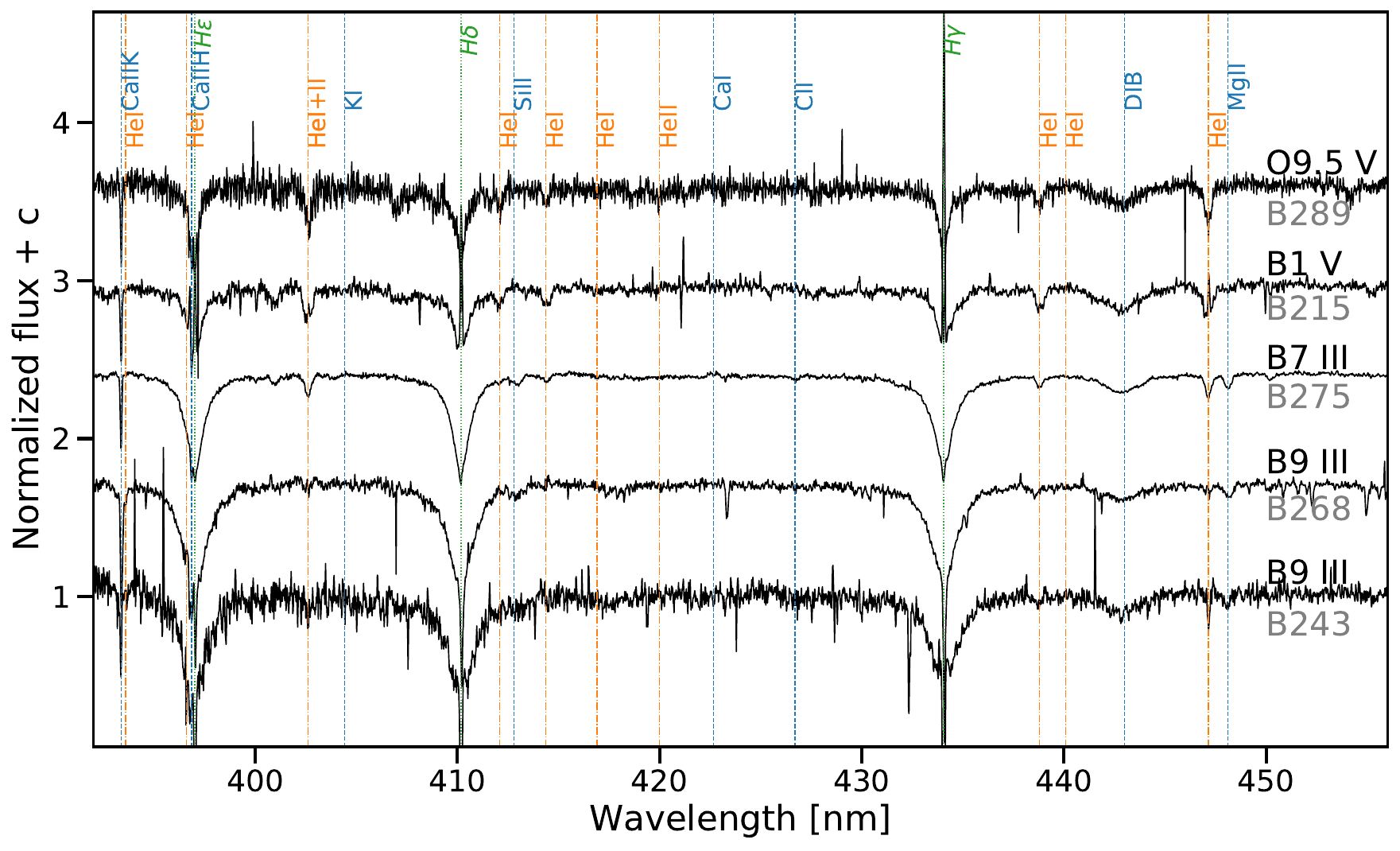}
      \caption{Blue X-Shooter spectrum of the PMS stars. The spectrum is stacked over all epochs and shows Balmer lines (green labels) and He\,{\sc I} lines (orange labels). Remaining features are marked in blue. B337 is not detected in the blue due to the high extinction. The spectra of B215 and B289 were not nebular subtracted, a sharp emission feature is present on top of the hydrogen features. Notable is the changing ratio of the Mg\,{\sc ii} 448.1\,nm (blue label) and He\,{\sc i} 447.1\,nm line strength with spectral type. A strong diffuse interstellar band (DIB) is present at 443\,nm (blue label). 
              }
         \label{classification}
\end{figure*}

\begin{figure*}
   \centering
   \includegraphics[width=\textwidth]{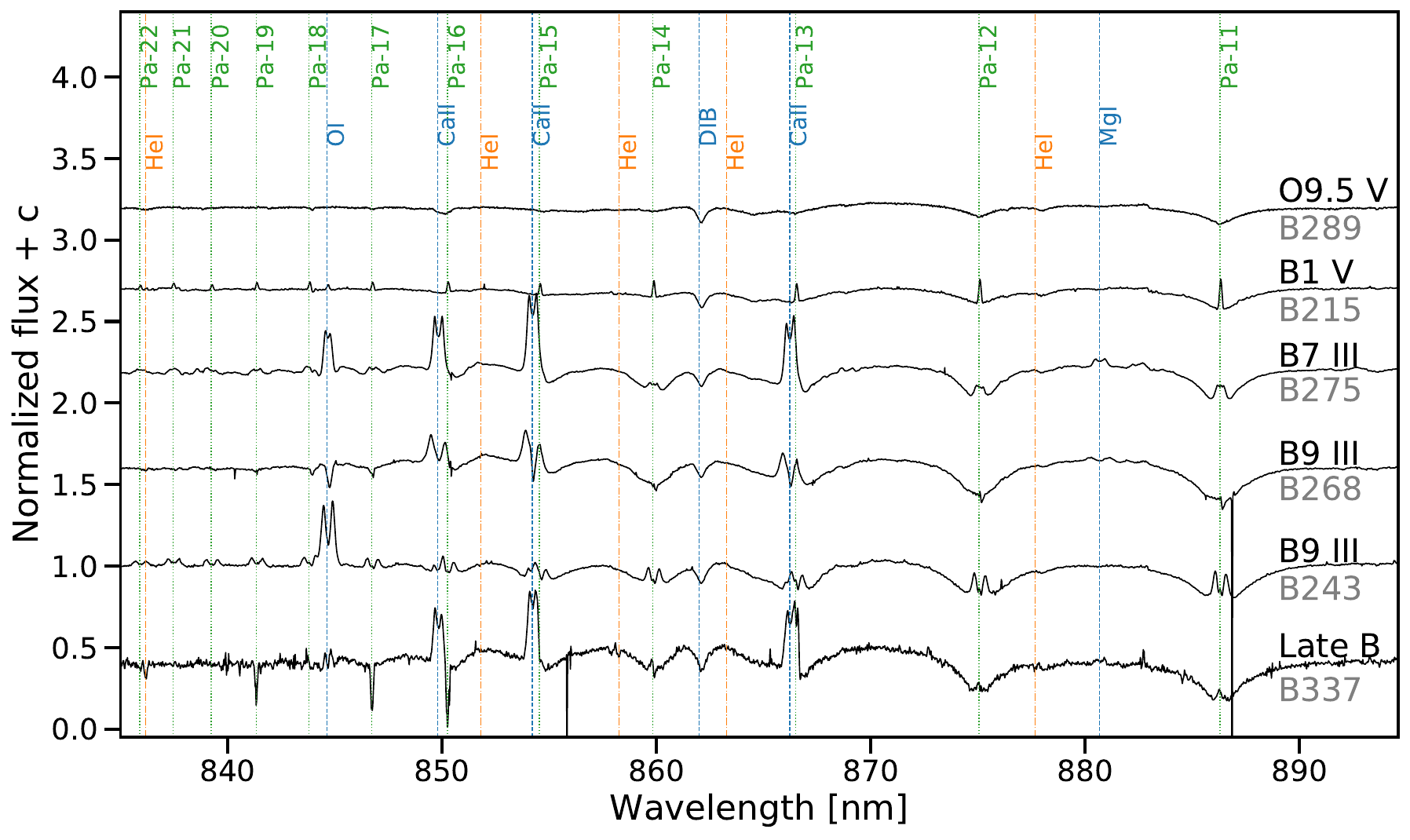}
      \caption{Part of the X-Shooter stacked spectrum centered at the Ca\,{\sc II} triplet. The spectra show broad photospheric profiles and the PMS stars with gaseous disks include double-peaked emission features. The spikes are residuals from the nebular emission subtraction. Note the Ca\,{\sc ii} triplet (blue labels) and O\,{\sc i} (blue label) emission in the Paschen series (green labels) lines. The Ca\,{\sc ii} triplet blends with Pa-13, Pa-15 and Pa-16.
              }
         \label{calcium}
\end{figure*}

\section{Updating the spectral classification}
Earlier classifications based on NIR spectroscopy by \cite{Hanson1997} and single epoch X-shooter observations by \RT\ can be improved by stacking the multiple spectra in this paper (we measured no radial-velocity shifts between the epochs, see Sec.~\ref{rv_shifts}, nor significant variability in the lines used for the classification). The higher signal-to-noise ratio together with the improved nebular subtraction allow for the detection of weak features that enable a detailed spectral classification (Figs.~\ref{classification} and \ref{calcium}).

\subsection{B215}
B215, previously classified as B0-B1\,V (\RT), shows a lack of He\,{\sc ii} lines and no C\,{\sc ii} at 426.7\,nm in the stacked spectrum, implying a B1 spectral type. This is further confirmed by the about equal strength of the He\,{\sc i} lines 414.4 and 412.1\,nm. Since there is no detection of O\,{\sc ii} at 435.0\,nm, the luminosity class V is confirmed \citep{Liu2019}, resulting in a B1\,V classification.

\subsection{B243}
B243 was previously classified as B8\,V (\RT) based on the He\,{\sc i} lines (402.6, 400.9 and 447.1\,nm). However, the nebular corrected and stacked optical spectrum of all epochs reveal weaker He\,{\sc i} lines than identified before. Since the Mg\,{\sc ii} 448.1\,nm line is stronger than the He\,{\sc i} 447.1\,nm, this star is classified as a B9 star. The luminosity class is taken as III because of a similar strength in He\,{\sc i} 402.6\,nm and Si\,{\sc ii} 412.8-413.0\,nm \citep{Liu2019}, implying a B9\,III.

\subsection{B268}
Earlier classifications of B268 lead to B2V or B9-A0 \citep[][\RT]{Hoffmeister2008}. The stacked spectrum of B268 reveals relatively strong Mg\,{\sc ii} 448.1\,nm lines compared to the neighboring He\,{\sc i} 447.1\,nm denoting a late-B/early-A type star. Additionally, since the He\,{\sc i} 402.6\,nm and 438.7\,nm lines are weakly present, this points to a B9 star. Finally, the star is classified as B9\,III, because of the comparable strength between He\,{\sc i} 402.6 nm and Si\,{\sc ii} 412.8-413.0\,nm \citep{Liu2019}. 

\subsection{B275}
The B7 III classification of B275 by \cite{Ochsendorf2018} is confirmed based on a relatively strong Mg\,{\sc ii} feature at 448.1\,nm compared to the He\,{\sc i} at 447.1\,nm with a ratio 3:2, indicating spectral type B7. The luminosity class is III due to the similar strength of the Si\,{\sc ii} lines at 412.8-3.0\,nm and \hea\ 402.6\,nm.

\subsection{B289}
This star shows He\,{\sc ii} absorption, but the He\,{\sc i} 447.1\,nm line is stronger than He\,{\sc ii} 454.0\,nm, so it must be an O8-O9 star. This is in line with the earlier classification of O9.7\,V by \RT. The He\,{\sc ii} 420.0\,nm absorption is visible in the stacked spectrum, it is weak and thereby pointing to O9-O9.7. Since He\,{\sc i} 438.8\,nm is stronger than He\,{\sc ii} 454.2\,nm and He\,{\sc ii} 468.6\,nm is stronger than C\,{\sc iii} 464.7/5.0/5.1\,nm, the spectral type is O9.5. Si\,{\sc ii} 411.6\,nm is weaker than He\,{\sc i} 412.1\,nm \citep{sota2011}, so the classification becomes O9.5\,V.

\subsection{B337}
The classification late-B by \RT\ is adopted for the most embedded star B337, based on SED fitting by \cite{Hanson1997}. Since stacking the spectrum did not improve the spectrum between 360-500\,nm enough to allow for a detailed spectral classification.

\section{Methods}
\label{sec:methods}
\begin{figure}
\centering
   \resizebox{\hsize}{!}
            {\includegraphics[width=\textwidth]{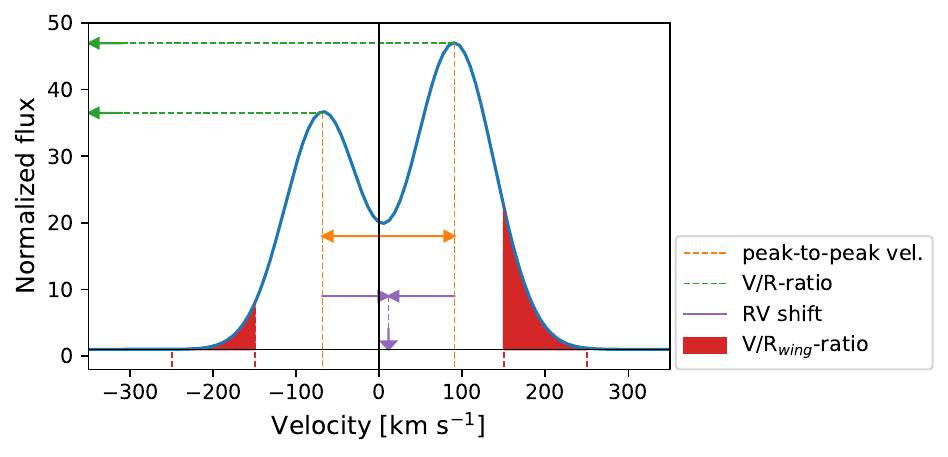}}
      \caption{A diagram showing the four diagnostics used to infer (variability) properties of double-peaked emission lines. Orange: the peak-to-peak velocity corresponds to the velocity difference between the central velocities of the two peaks. Green: the heights of the peaks are divided to obtain the V/R-ratio. Purple: the difference between the middle of the central peaks and the rest wavelength gives a central wavelength shift or radial-velocity shift. Red: the equivalent width of this line is determined between -250 and -150\,km\,s$^{-1}$ and between 150 and 250\,km\,s$^{-1}$. The (V/R)$_{\rm wing}$-ratio is the ratio of these values.
              }
         \label{explainvr}
   \end{figure}

In this section we discuss the methods used to identify and characterize lines originating from the circumstellar material, to determine spectral properties of double-peaked emission lines and to measure the temporal variability of circumstellar lines. The method section is similarly structured as the subsequent sections where we present the results (Secs.~\ref{sec:circumstellar}-\ref{sec:varia}). 

\subsection{Circumstellar lines}\label{sec:method_circumstellar}
\RT\ identified emission lines that they link to a circumstellar rotating disk; these lines are often  double peaked and rest on top of a photospheric absorption line. We identify these and more circumstellar features by subtracting a stellar model from the spectrum. The stellar model is a BT-NextGen atmospheric model computed with the PHOENIX code \citep{hauschildt1999, starmodels2012RSPTA.370.2765A} with stellar parameters as determined by \RT\ with FASTWIND modeling. The residual spectrum, referred to as the disk spectrum, shows the circumstellar material, which can have different components; a disk, disk wind, accretion flow and/or jet. They provide the opportunity to study the structure, the dominant physical processes and the material in the disk.

Additionally, the EW (and its errors) for disk emission lines is calculated following \cite{Vollmann2006} for low- and high-flux lines. For a handful of emission lines in each star we determine the EW, since we only want to measure uncontaminated lines, which is only possible for lines with little to no nebular subtraction residuals, therefore we exclude B337 from this procedure. The measured disk lines and the corresponding EWs are listed in Tab.~\ref{tab:EW} for B243, B268 and B275. 

\subsection{Double-peaked line properties} \label{sec:line_properties}
The double-peaked emission lines are well pronounced and the two peaks can be modeled with a double Gaussian profile and a linear component, the latter to model the local continuum. Fig.~\ref{explainvr} shows a hypothetical disk-only spectral line that would result from star model subtraction. We illustrate four diagnostics used to infer characteristic line properties. Two of them are discussed in this section: a) peak-to-peak velocity and b) V/R-ratio. The other two diagnostics are discussed in Sec.~\ref{method:radvel} and \ref{sec:temporal_line_prop} since they are evaluated over time.

The peak-to-peak velocity is measured as the distance between the center of the two peaks of the Gaussian model, illustrated by the orange arrows in Fig.~\ref{explainvr}. Half of this velocity yields an approximation for the projected rotational velocity of the disk region where the bulk of the line emission originates.

The heights of the blue (V) and red (R) peak of the Gaussian model are obtained after correcting the peak fluxes for the linear continuum, illustrated by the green arrows in Fig.~\ref{explainvr}. The ratio between these two peak values is the V/R-ratio. The ratio is unity in case of an identical peak height. 

\subsection{Spectroscopic variability}\label{sec:methods_temp}

\subsubsection{Temporal variance spectrum (TVS)} \label{sec:tvs_method}

Spectroscopic variability is identified in spectral lines using the temporal variance method \citep[TVS;][]{Fullerton1996,dejong2001}. This method compares the observed standard deviation of the flux $\sigma_{\rm obs}$ with the expected standard deviation of the flux $\sigma_{\rm exp}$ in a given wavelength bin. If the ratio of the two standard deviations is above unity, the variation in the spectral line between the epochs is larger than what can be expected from the noise at that particular wavelength. When no variation is detected, the value of the TVS is randomly distributed around unity due to the noise in the data.

The $\sigma_{\rm exp}$ is calculated by averaging the standard deviation in the continuum regions around the spectral line in each epoch, yielding $\sigma_{\rm con}$, which is scaled by the square root of the ratio between the mean continuum flux and the mean (normalized) flux of the evaluated wavelength bin $j$ for all epochs of observation. The $\sigma_{{\rm obs},j}$ is directly calculated from observed fluxes in the wavelength bin. The ratio is given by, 
\begin{eqnarray}
{\rm TVS}_{j} = \frac{\sigma_{\rm obs, j}}{\sigma_{\rm exp, j}} = \frac{\sigma_j}{\sigma_{\rm con}} \sqrt{\frac{\bar{F}_{\rm con}}{\bar{F}_j}}, 
\end{eqnarray} where $\bar{F}_{con}$ is the average continuum flux and $\bar{F}_j$ the average flux in the wavelength bin $j$ over all epochs. To determine the significance of the variability, the TVS value is compared to a $\chi^2$-distribution where the degrees of freedom are given by the number of observations plus the amount of lines that have a similar TVS pattern. Only variability at a probability limit of 95\% is considered. 

In this study, we apply the TVS method to detect the line-of-sight velocity regimes which display significant variability. Assuming that the variability originates from the circumstellar disk and that the disks are in Keplerian rotation, we calculate where these velocities would originate in a rotating disk by assuming an inclination angle of the disk relative to the line-of-sight \citep[based on earlier studies of these stars][]{Backs2023, poorta2023}. The TVS method does not produce realistic results in the central parts of lines that have a nebular contribution, as residual nebular emission features between observing epochs cause a spurious variability signal. 
Inspection of the TVS signal generally provides a good way to discriminate between a spurious variability signal, which is often restricted to a few velocity bins, and intrinsic line variability over a wider velocity range or seen in multiple lines \citep{Fullerton1996}. 

Anticipating on the results of this study, we find that the TVS method indeed discriminates between sources that display significant line variability and sources that do not. B243, B268, and B275 are in the former category and B215, B289 and B337 do not show detectable variability (see Fig.~\ref{B215_tvs}, \ref{B289_tvs} and \ref{B337_tvs}). In particular for B337, the spectra are relatively noisy and nebular contamination is severe, making the TVS signal too weak to allow for any conclusions to be drawn, the TVS may show a marginal detection at most (in the blue peak of the Ca\,{\sc ii} triplet, see Fig.~\ref{B337_tvs}).

\subsubsection{Radial-velocity shift}\label{method:radvel}

The radial velocity of a system is measured as the difference between the central velocity of a photospheric absorption profile and the rest wavelength of the spectral line. However, one may also derive the radial velocity from double-peaked disk lines (if the disk is relatively symmetric and not circumbinary) by taking the difference between the middle of the blue and red peak and subtracting the rest wavelength value, which is the method employed in this study. The striped purple arrow in Fig.~\ref{explainvr} points to the middle velocity of the blue and red peak. 

\begin{figure*}[h]
   \centering
   \includegraphics[width=0.95\textwidth]{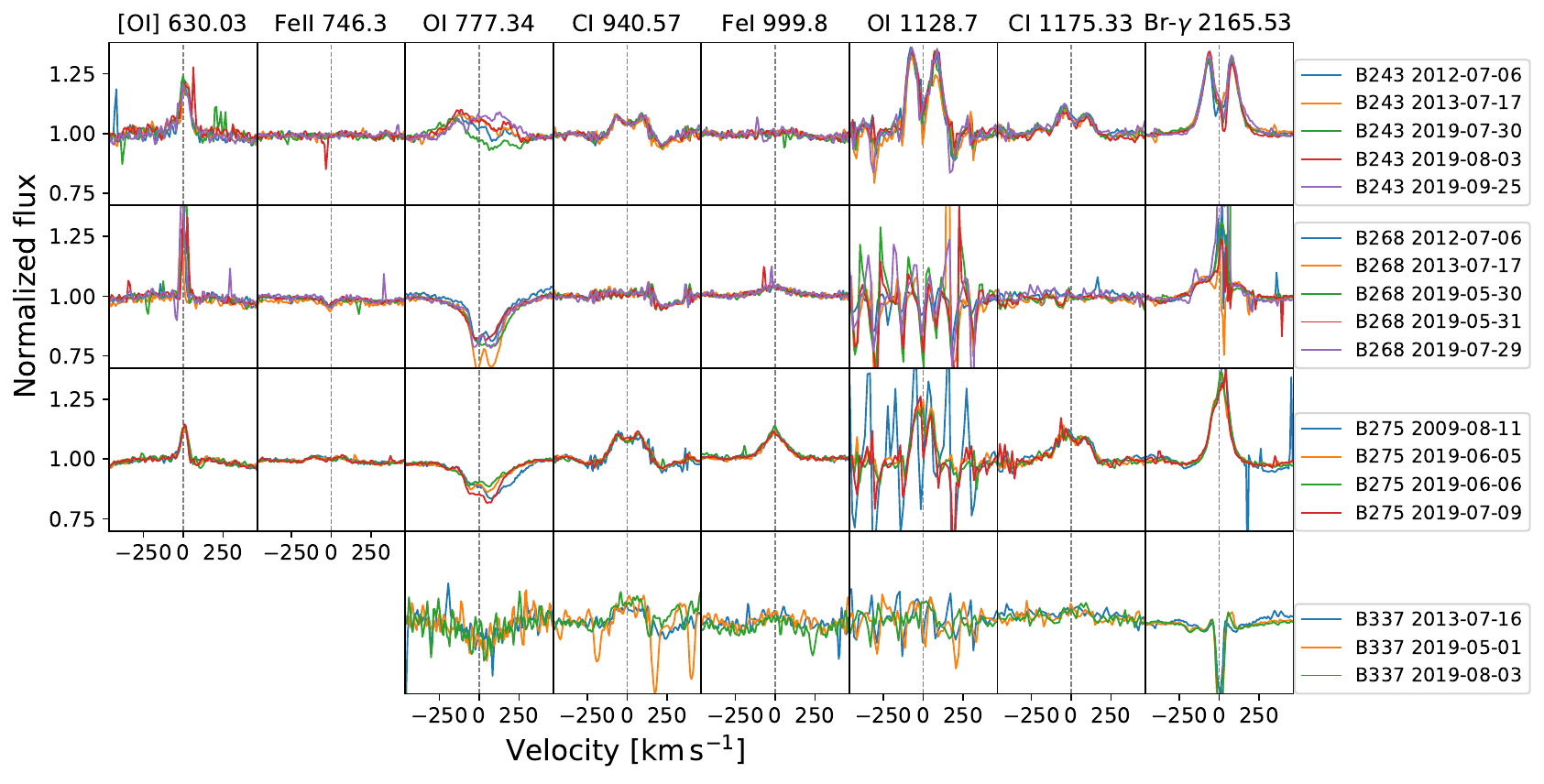}
      \caption{Overview of O\,{\sc i}, [O\,{\sc i}], Fe\,{\sc i}, Fe\,{\sc ii}, C\,{\sc i} and Br-$\gamma$ lines in B243, B268, B275 and B337. The multiple epochs are over-plotted and the wavelengths are shown above each panel in nm. The [O\,{\sc i}] and Fe\,{\sc ii} line of B337 are not shown, since the bluer part of the X-Shooter spectrum is too noisy. All stars display some nebular emission (residuals) on the Br-$\gamma$ emission (right panel). The variability between the epochs is discussed in Sec.~\ref{missing_lines}. 
              }
         \label{missing_lines}
\end{figure*}

\subsubsection{Variations in double-peaked line properties} \label{sec:temporal_line_prop}
We calculate the ratio of the EW in corresponding velocity intervals in the blue and red wings of the line, (V/R)$_{wing}$-ratio, to avoid nebular contamination (not depicted in Fig.~\ref{explainvr}) and investigate the innermost parts of the disk. In Fig.~\ref{explainvr} this is illustrated for the velocity intervals [$\pm150$,$\pm250$]\,km\,s$^{-1}$. Wherever this property is used throughout the analysis, the velocity interval is clearly indicated in the relevant figures. The wavelength ranges are chosen to enclose the largest interval without contamination in any of the lines. The chosen velocity regimes are the same for every line and epoch of a target. 

The (V/R)$_{\rm wing}$-ratios, peak-to-peak velocities and V/R-ratios are measured for each epoch in order to search for trends in their temporal variations. 

\begin{figure}[h]
   \centering
   \includegraphics[width=\hsize]{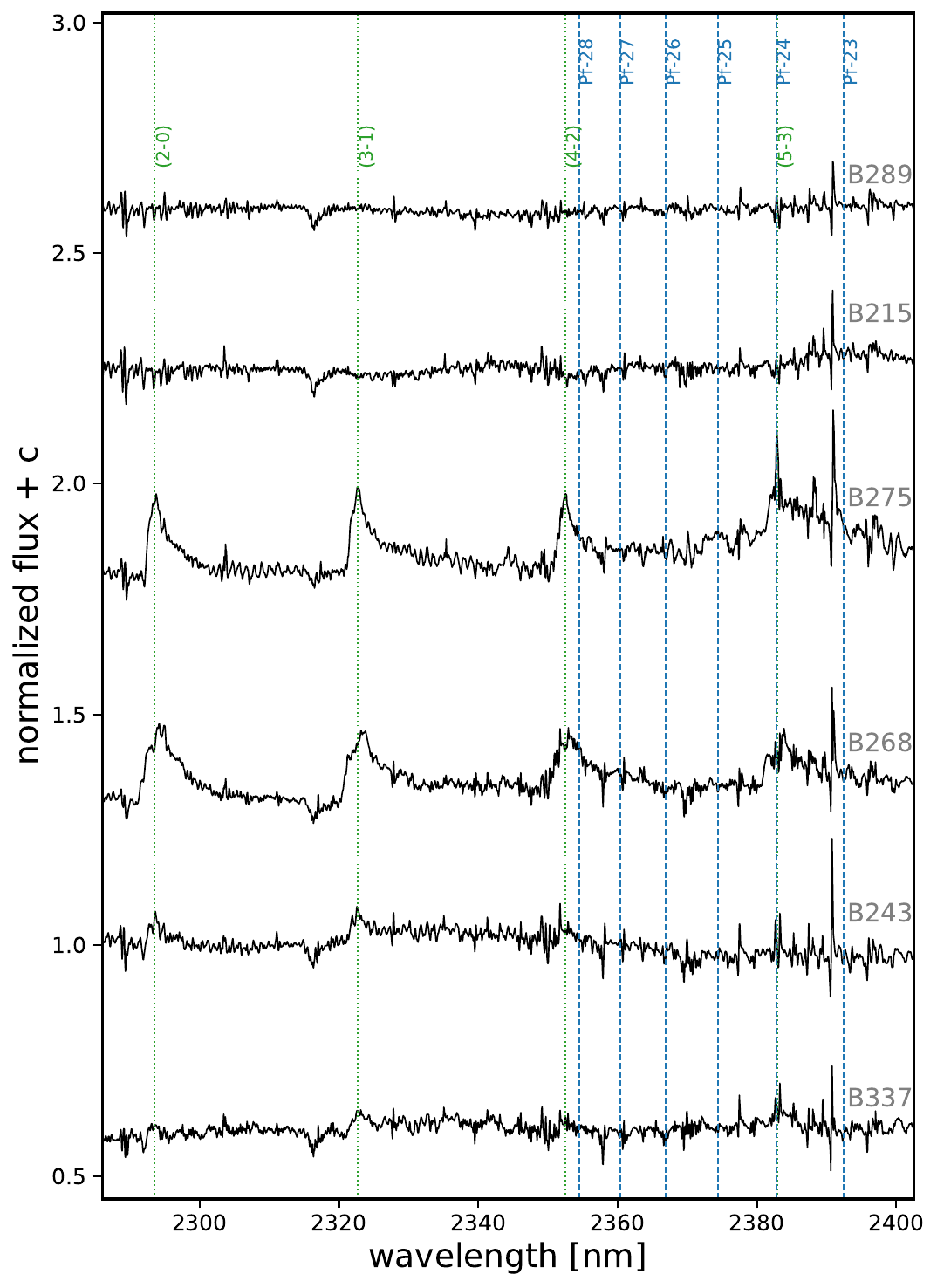}
      \caption{Stacked NIR X-Shooter spectra of the PMS stars. All sources apart from B215 and B289 display CO bandhead emission (green dotted vertical lines), indicating the presence of a hot circumstellar disk. We detect CO in B337 for the first time. The blue dashed lines show the Pfund series. The narrow absorption features correspond to residuals from the telluric correction.
              }
         \label{cobandheads}
\end{figure}

\section{Circumstellar lines}\label{sec:circumstellar}

In this section we describe the lines in the disk spectrum (after stacking and stellar model subtraction, see Sec.\,\ref{sec:method_circumstellar}) of the four stars with gaseous disks (B243, B275, B268 and B337).  

The stacked spectrum allows for a detailed characterization of lines; no (inverse) P\,Cygni profiles are identified and the stars show (newly detected for B337) CO bandhead emission (Fig.~\ref{calcium}-\ref{cobandheads}). Other features in the disk spectrum are lines from H\,{\sc i}, O\,{\sc i} and Ca\,{\sc ii} in all targets, and Na\,{\sc i}, Mg\,{\sc i}, C\,{\sc i}, Fe\,{\sc i}/Fe\,{\sc ii} and [O\,{\sc i}] in some.

A detailed discussion on the presence, strength and shape of the spectral lines is provided in App.~\ref{app:spectral_lines}. A characterization of the lines in the circumstellar spectra of the stars is given in the next section, supported by Tab.~\ref{tab:emission_line}.
The table consists of emission and absorption lines that show in the spectrum. For the latter that means they are not present in the star model and could not be tight to a stellar origin (see Sec.~\ref{sec:source_varia_diff} for a discussion on this). Despite the nebular subtraction described in Sec.~2, nebular line residuals remain present in some of the lines, especially in H\,{\sc i}. Consequently, the central several tens of km\,s$^{-1}$ of these lines cannot be used. 

All stars show H\,{\sc i} emission, the strongest double-peaked Balmer, Paschen and Bracket emission is present in B243 and the weakest in B337 (Fig.\,\ref{calcium}). 

O\,{\sc i} lines are observed in all stars in absorption and emission. However, only B243, B275 and B337 have double-peaked emission at 844.6\,nm and 1128.7\,nm. They are strongest in B243, which is the only one with the O\,{\sc i} triplet at 777.4\,nm in emission. 

The Ca\,{\sc ii} triplet is weakest in B243, weaker than the overlapping double-peaked Paschen features. In the other objects, these lines are stronger than the Paschen features. In particular, the peaks in B275 and B337, are stronger and closer together than the lines in B268. 
A similar trend is seen for Mg\,{\sc i} double-peaked emission at 880.7\,nm. This line is strongest for B275, weaker for B268, weakest in B337 and not present in B243.

Comparable behavior is seen in Fe\,{\sc ii} at 999.8\,nm; strongest in B275, absent in B243. 
Other Fe\,{\sc i} and Fe\,{\sc ii} lines appear in the spectra as weak absorption or double-peaked emission in B268 and B275. The stacked disk spectrum of B337 has a too low S/N to detect the weak Fe lines. 

Multiple double-peaked C\,{\sc i} emission is present in the spectra of B243 and B275, the two stars where O\,{\sc i} 844.6\,nm emission is observed and with the strongest hydrogen lines. The carbon lines are detected at wavelengths > 900\,nm. In B275, these lines are mostly stronger than the Fe double-peaked emission lines. One C\,{\sc i} line is also present in B337, which shows a small O\,{\sc i} 844.6\,nm emission line. 

B268 is the only star with Na\,{\sc i} (589.0\,nm and 589.6\,nm) absorption. These absorption lines reside alongside narrow absorption lines originating from the interstellar medium. 

Lastly, the CO bandheads are strongest in B268 and B275, and much weaker in B243 and B337 (Fig.~\ref{cobandheads}). This is the first detection of CO bandhead emission in B337. The strength of this feature depends also on the continuum emission \citep{poorta2023}. 

\begin{table}
\caption{Circumstellar spectrum overview in B243, B268, B275 and B337.}
\centering
\begin{tabular}{l|llll|l}
\hline\\[-8pt]
Circumstellar lines & B243 & B268 & B275 & B337 & Fig. \\ 
\hline
\hline \\[-8pt]
Higher order H\,{\sc i} lines &   D$^{\rm V}$  &     &  D$^{\rm V}$ &   & \ref{calcium}-\ref{missing_lines} \\
Lower order H\,{\sc i} lines &  D$^{\rm V}$  &  U$^{\rm V}$   &  U$^{\rm V}$ &  U & \ref{calcium} \\
Na\,{\sc i} doublet   &     &   A$^{\rm V}$  &   &   & \ref{B268_TVS} \\
$[$O\,{\sc i}$]$ 630.0\,nm  &  S  &   S  & S &   &  \ref{missing_lines} \\
Fe\,{\sc i} \& Fe\,{\sc ii} lines &    &  A  &  D &    & \ref{missing_lines}\\
O\,{\sc i} 777.3\,nm    & C$^{\rm V}$   &   A$^{\rm V}$   &  A$^{\rm V}$  &  A   & \ref{missing_lines} \\
O\,{\sc i} 844.6 \& 1128.7\,nm & D$^{\rm V}$   &     & D$^{\rm V}$  &  (D)  & \ref{calcium}-\ref{missing_lines}\\
C\,{\sc i} lines  & D    &      & D &   D  & \ref{missing_lines} \\
Ca\,{\sc ii} triplet  &  (U)    & D$^{\rm V}$    & D$^{\rm V}$ &  D$^{\rm V}$  & \ref{calcium} \\
Mg\,{\sc i}  &      & D   & D &  D & \ref{calcium} \\
O\,{\sc i} 926.2-926.6\,nm    & A$^{\rm V}$   &   A   &  A  &  A   & \ref{B243_TVS} \\
Fe\,{\sc i} 999.8\,nm  &      & S    &S  &  S  & \ref{missing_lines}\\
CO bandheads &  (C)   &  C  &  C &  C  & \ref{cobandheads} \\
\hline
\end{tabular}
\begin{tablenotes}
    \small
    \item \textbf{Notes.} The presence and characteristic of the emission lines is indicated with 'D' if the line is double-peaked, 'S' for a single-peaked line, 'C' when the emission is too complex to recognize a single or double peak, 'U' when it is unclear whether the line is double peaked or single peaked due to nebular subtraction residuals and 'A' if the line is observed in absorption. Brackets mean that the line detection is weak and $^{\rm V}$ is added for a variable line, which is explored in Sec.~\ref{sec:varia}. Not all stars show the same lines in their spectrum. Largest differences in occurrence are between B243 and B268. Most emission lines are present and pronounced in B275. The last column indicates in which figure(s) the lines can be inspected.
\end{tablenotes}

\label{tab:emission_line}
\end{table}

\begin{figure}
   \centering
   \includegraphics[width=\hsize]{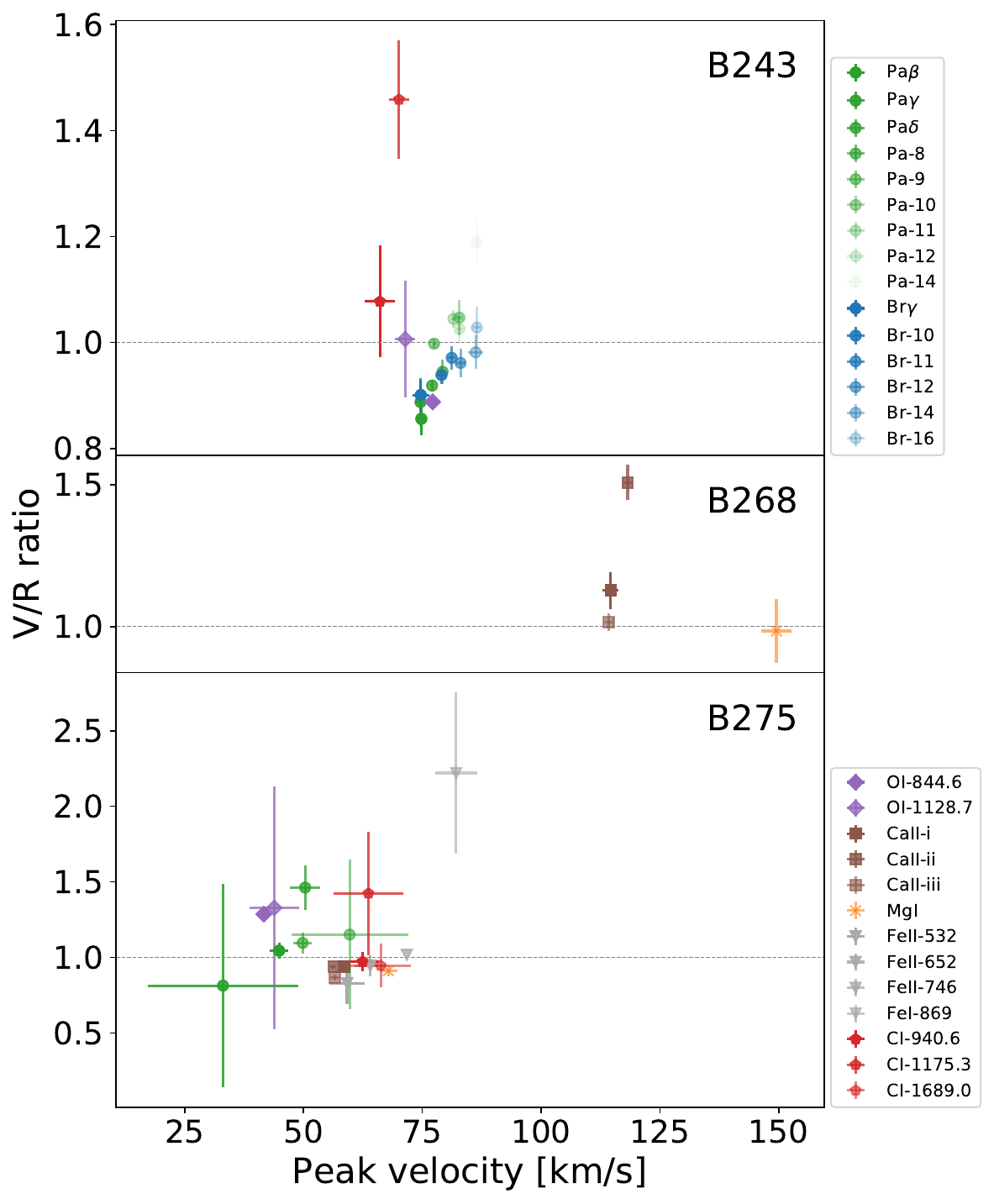}
      \caption{The peak velocity (half the peak-to-peak velocity) of all double-peaked emission lines in B243, B268 and B275 is plotted against their V/R ratios. The data points represent the line properties in one epoch of each star. The emission lines in a single epoch of a star have different V/R ratios and peak velocities. For B243 and B275, the lower excitation lines form at relatively lower velocities, hence at more slowly rotating parts of the disk and their V/R ratios are relatively smaller, meaning their red peak is stronger than the blue, than measured for the higher excitation lines. For both stars this results in a trend of a higher V/R ratio with a higher disk velocity. The displayed epochs are for B243: 2012-07-06, B268: 2019-05-31 and B275: 2019-07-09. }
         \label{diskvsosc}
\end{figure}

\section{Double-peaked line properties}
In this section we discuss the results from the line diagnostics described in Sec.~\ref{sec:line_properties}. B337 is not shown in this section due to the lack of double-peaked emission lines. 

\subsection{Peak-to-peak velocities} \label{sec:peakvel}

The peak-to-peak separation ranges between 135-180\,km\,s$^{-1}$ for B243, 220-300\,km\,s$^{-1}$ in B268 and 80-160\,km\,s$^{-1}$ for B275
. Half of the peak-to-peak velocities, the so called disk velocities, for one epoch per star are shown on the x-axis of Fig.~\ref{diskvsosc}.

In Fig.~\ref{diskvsosc} B243 shows a trend where the projected disk velocity is higher for higher-order hydrogen lines (indicated by the more transparent colors). This can be understood in terms of disk rotation and that the lower excitation lines (indicated by the darker colors) are formed over a larger extent of the disk up to where the velocities are lower with a correspondingly larger radiation area.
This trend is observed in the hydrogen (green and blue), oxygen (purple) and carbon lines (red). This behavior is similarly observed in the hydrogen lines of B275 (green), while other lines are more scattered. High projected disk velocities are measured for the Ca\,{\sc ii} triplet, Mg\,{\sc i} and C\,{\sc i} lines
. The velocities of the O\,{\sc i} lines in B243 and B275 are tens of km\,s$^{-1}$ lower, more similar to the lower excitation hydrogen lines. 

   \begin{figure*}
   \centering
   \includegraphics[width=\hsize]{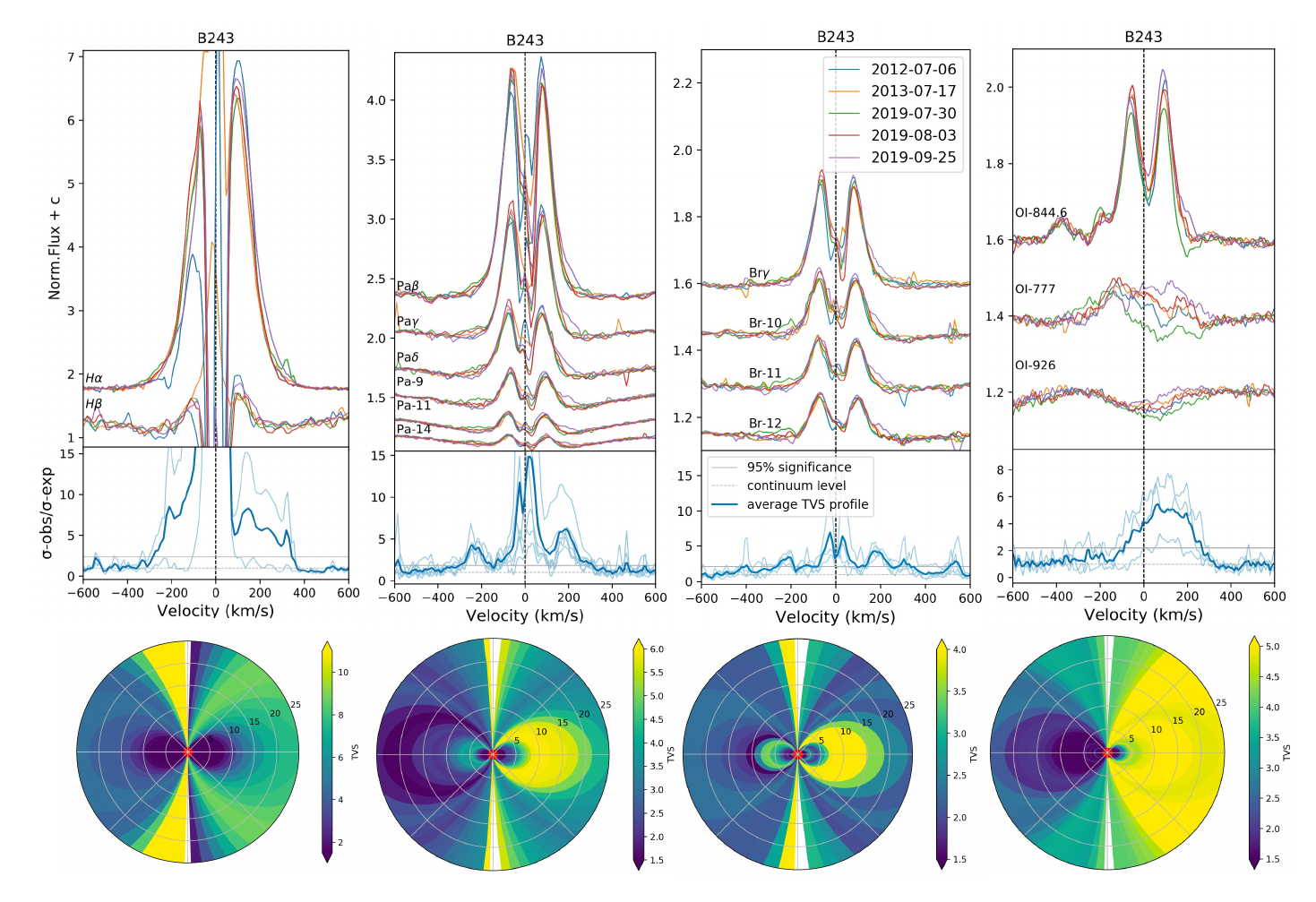}
      \caption{TVS and diskmaps for B243. The emission lines in B243 that display variability are shown in the upper panel (first row). The middle panel displays the TVS for each emission line (light blue) and an average profile (dark blue) calculated for each series of lines displayed in the upper panel. The variations at the line center of the H\,{\sc i} lines are residuals from the nebular subtraction. The TVS panel shows the continuum level as a dotted line at unity and a gray line marking the level at which variations reach a 95\% significance level. The average TVS in the lower panels in each column are mapped on to the circles in the bottom row. They show a face on disk view, where the TVS is mapped according to their velocity bins. The gray circles and accompanying numbers are in units of the stellar radius. 
      The white areas are the result of the disk projection and do not have any physical meaning.
              }
         \label{B243_TVS}
   \end{figure*}
   
\subsection{V/R - ratio}\label{sec:VR}
Fig.~\ref{diskvsosc} shows the V/R-ratio in all stars for double peaked lines. Remarkably in B243 and B275 the values vary from lower to higher than unity for transitions in a single epoch, meaning that the highest peak is sometimes the red and sometimes the blue one. A common interpretation is that the V/R-ratio traces an asymmetry of the circumstellar disk. The observations suggest a higher degree of complexity (azimuthal variations like a spiral arm, see Sec.\,\ref{dis:peaktopeak_vs_VR-ratio}). 
B243 and the H\,{\sc i} lines in B275 both show a trend where the lines with the lowest V/R-ratio are the lines with the lowest peak velocity.

   \begin{figure*}
   \centering
   \includegraphics[width=0.84\hsize]{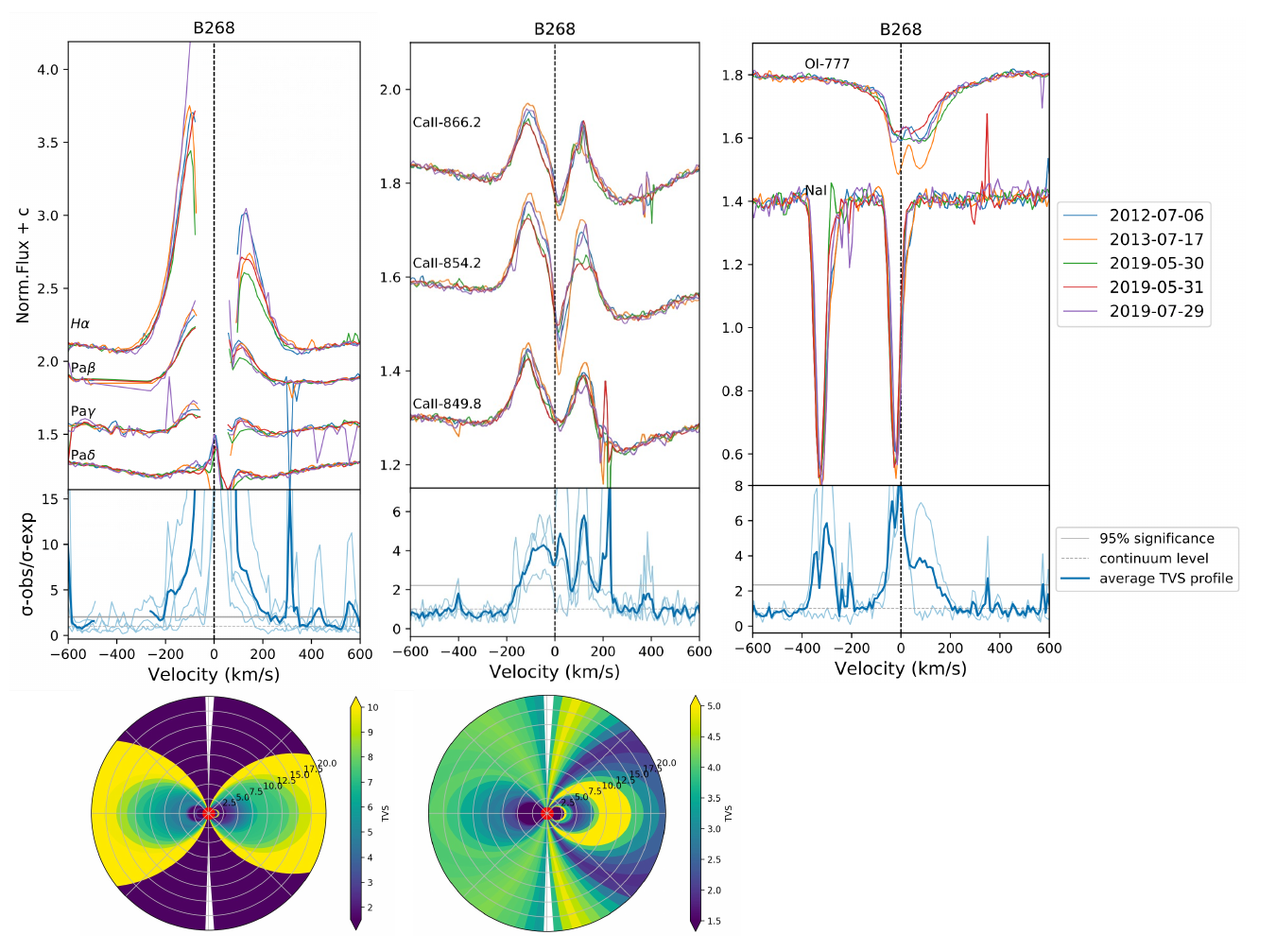}
      \caption{TVS and diskmaps for B268, see caption Fig.~\ref{B243_TVS}.
      Despite the blue emission peaks being stronger, variability seems to be relatively equal on both sides. 
              }
         \label{B268_TVS}
   \end{figure*}

\section{Spectroscopic variability}\label{sec:varia}
In this section we discuss the results from characterizing the variability with the TVS method, potential radial-velocity changes and temporal variance in the line diagnostics (Sec.~\ref{sec:methods_temp}) to quantify the variability of the PMS stars.

\subsection{Temporal variance spectrum (TVS)}
\label{P1:sec:TVS}

The cadence of our observations allows to probe variations on timescales of days, months and years. In principle, one could use the individual nodding positions, taken at intervals of 10$-$20\,min, to also probe characteristic timescales of about an hour. On these timescales, however, no intrinsic variability is seen. 

We find that the TVS-method reveals significant variations in the H\,{\sc i}, O\,{\sc i}, Ca\,{\sc ii} and Na\,{\sc i} lines of B243, B268, and B275 up to velocities of 
$\sim$360, $\sim$320, and $\sim$370\,km\,s$^{-1}$ respectively, in both the red and blue wings of the lines. These velocities are significantly higher than the observed v\,sin$i$ (see Sec.\,\ref{dis:diskvel}). 
H$\alpha$ shows the strongest variations on the shortest ($\sim$ days) timescales. Lower order hydrogen lines show more pronounced variability than their higher order counterparts.

Generally, the velocity interval in which variability is observed is consistent across line series, e.g., the Paschen or Bracket series. The velocity interval and amplitude of the variability differ for different atomic species in a single source (e.g. in B243 the O\,{\sc i} lines vary at lower velocities on the blue side than seen in the H\,{\sc i} lines). Also, the variability detected in one atomic species may differ between the different stars (e.g. the Paschen series in B243 shows variability on the red and blue side, but the Paschen series in B275 shows mostly variability on the red side). However, for all stars the O\,{\sc i} lines show stronger variability at the red side of the lines compared to the blue side.

\begin{figure*}
   \centering
   \includegraphics[width=\hsize]{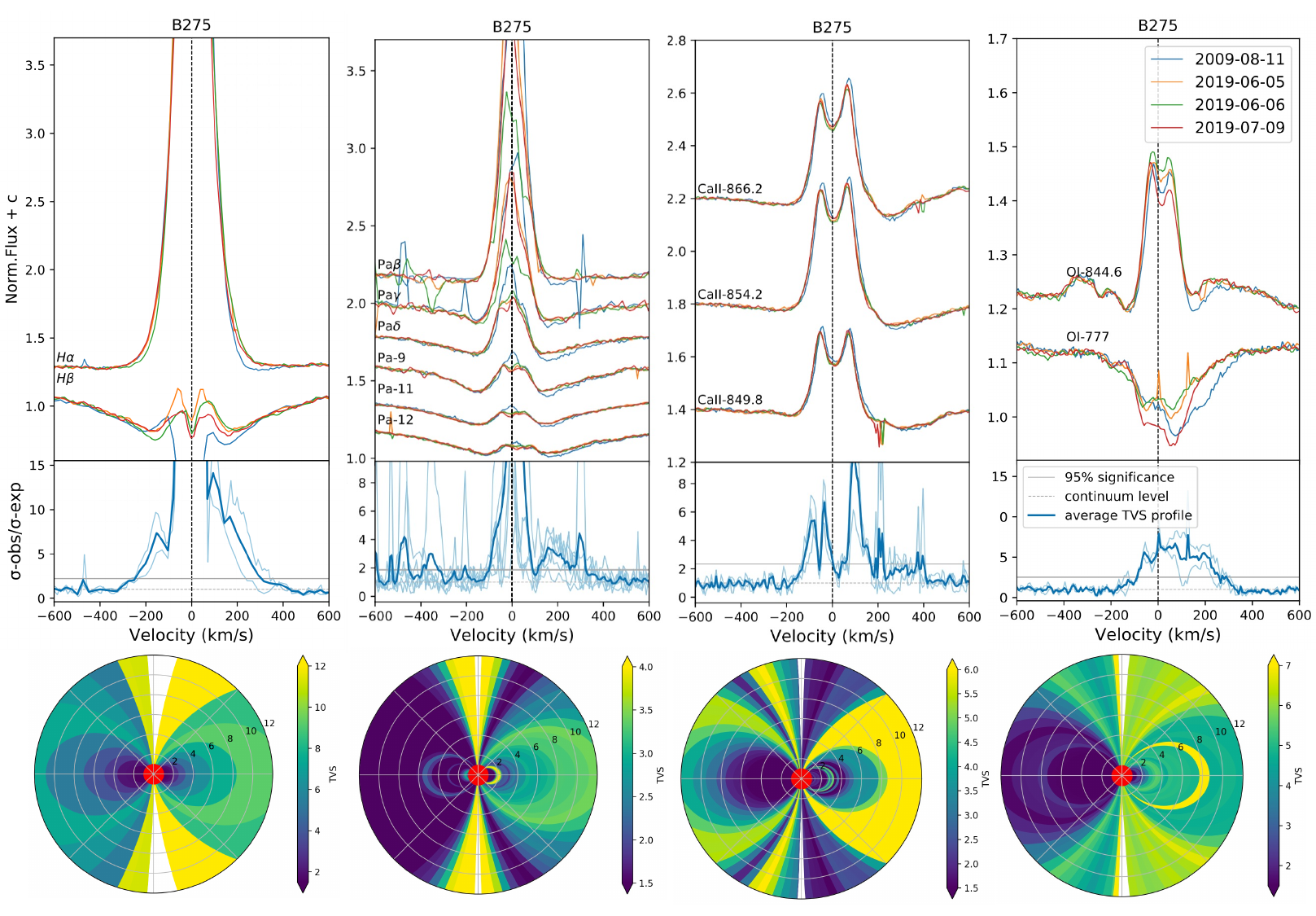}
      \caption{TVS and diskmaps for B275, see caption Fig.~\ref{B243_TVS}.
      In all lines, the red side shows more variability close to the star than the blue side.  
              }
         \label{B275_TVS}
   \end{figure*}

To help visualize the location in the disk where the variability originates, the 1D TVS profile is mapped to a 2D Keplerian disk model (see Sec.\,\ref{sec:tvs_method}). 
The colors in the disk maps display the average TVS profile as measured for a series of lines in a certain object, and are shown in the middle panels of Fig.~\ref{B243_TVS}-\ref{B275_TVS}. The height of the color bar is determined for each average TVS profile based on the highest TVS measure outside of the nebular line interval, which is not shown in the map. This leaves a blue color on the regions with a zero velocity with respect to the observer. 
Each velocity bin of the TVS profile is projected on a face-on view of the disk adopting an average inclination of 70$^{\circ}$ for the sources. Inclinations of 60$^{\circ}$ or 80$^{\circ}$ do not substantially affect the projected maps. The left side of the disk map shows the blue-shifted velocities and the right side of the disk map the red-shifted velocities. 

In the following parts we discuss the (significant) variability detected in the three PMS stars with a gaseous disk.
   
\subsubsection*{B243}
B243 displays significant variability in the hydrogen Balmer, Paschen and Bracket series, as well as in the O\,{\sc i} lines. We detect variability on the blue and red sides of the lines at velocities up to $\sim360$\,km\,s$^{-1}$ in H$\alpha$ and up to $\sim350$\,km\,s$^{-1}$ for the Paschen and Bracket series (Fig.~\ref{B243_TVS}). The O\,{\sc i} lines show strong variability on the red side at velocities up to $\sim300$\,km\,s$^{-1}$, and variability up to much lower velocities of $\sim180$\,km\,s$^{-1}$ on the blue side. The detected variations are similar for the O\,{\sc i} lines and the H\,{\sc i} lines, as shown in their (average) TVS profiles. 
However, these average H\,{\sc i} and O\,{\sc i} TVS profiles differ as seen on the disk maps, while the profiles of the Paschen and Bracket series are very similar to each other;
The central two disk maps in Fig.~\ref{B243_TVS} for the Paschen and Bracket series show a similar profile: variability on both sides of the disk and slightly stronger on the red-side than on the blue side. The leftmost and rightmost maps also show strong red side variability, but while hardly any variability is detected at the blue side of the O\,{\sc i} lines, very strong variability is detected on the blue side of the Balmer lines.

Since the Ca\,{\sc ii} triplet emission is weak compared to the imposed Paschen emission, this feature is excluded from the analysis. We detect variability in all lines for the shortest probed timescale, which is four days.

\subsubsection*{B268}
Variability in B268 is observed in the strongest hydrogen lines (up to Pa$\delta$), the Ca\,{\sc ii} triplet, O\,{\sc i} at 777\,nm triplet, and the Na\,{\sc i} doublet. H$\alpha$ and the Pa series are characterized by a strong blue peak and variability on both sides of the feature up to 320\,km\,s$^{-1}$ for the former and up to $220$\,km\,s$^{-1}$ for the latter (Fig.\,\ref{B268_TVS}). The red peak variations of the Ca\,{\sc ii} lines cannot be trusted, due to residuals from the subtraction of the overlapping nebular Paschen emission lines. All three calcium lines show similar variations in the blue peak up to 160\,km\,s$^{-1}$. This variability is similar to the variability in the blue peak of the Paschen series, where the first two observations in 2019 show the weakest features compared to other epochs. Furthermore, variations in the Na\,{\sc i} doublet are characterized by additional absorption superimposed on the red side of the interstellar Na\,{\sc i} contribution, which only shows in the observations of 2012 and 2013. Variations in the O\,{\sc i} triplet are detected on the red side and show between all epochs.

\subsubsection*{B275}
Variations in B275 are most pronounced on the red side of all emission lines, with the exception of the Ca\,{\sc ii} lines. Variability is detected at velocities up to $\sim370$\,km\,s$^{-1}$ for the Balmer and Paschen series (Fig.\,\ref{B275_TVS}) and up to $\sim320$\,km\,s$^{-1}$ for the O\,{\sc i} lines. All species show variability between the 2009 observation and those in 2019. 
Additionally, monthly and daily variability is detected in the lower order hydrogen lines and in the O\,{\sc i} lines.

In the double-peaked emission of \oi\ 844.6\,nm (at $\pm 120$\,km\,s$^{-1}$), the blue peak has a higher amplitude than the red one, but the variability in the red peak is stronger as shown in the TVS profile (Fig.\,\ref{B275_TVS}). In contrast, the Ca\,{\sc ii} triplet shows a stronger red peak, and overall, the 2009 spectrum displays stronger lines.

\subsection{Radial-velocity shifts}\label{rv_shifts} 
Earlier studies of massive young stars in M17 reported a lack of short period binaries, revealed by the low radial-velocity dispersion (\srv) observed in single epoch observations of a sample of 12 stars \citep{sana2017}. Our multi-epoch approach provides the opportunity to check for radial velocity variations (indicative of close binarity) in our sample stars.

There are no radial-velocity changes by measuring the disk lines >\,$20$\,km\,s$^{-1}$ from epoch to epoch for all of the stars, a limit often used for short-period-binary detection \citep{sana13}. Furthermore, the shifts in the double peaks correspond with the shifts in the radial velocity of the measured interstellar absorption lines, that we measured for reference. 
These shifts are therefore attributed to differences in wavelength calibration between the epochs. This is supported by the estimated radial-velocity shift of the interstellar absorption lines. This demonstrates the potential of using the center of the double peaks to measure radial-velocity shifts.

\subsection{Peak-to-peak velocity variations} \label{p1:sec:DP_emission}

Similar to what is observed for a single epoch (Fig.~\ref{diskvsosc}), the peak-to-peak velocities differ among the double-peaked lines and their respective temporal differences are generally similar.

Only small ($<20$\,km\,s$^{-1}$) peak-to-peak velocity variations are detected on timescales of days, months and years for all PMS stars (Fig. \ref{fig:peaktopeak}). The three epochs in 2019 in B243 show a marginal temporal trend in their respective variations of $\sim5$\,km\,s$^{-1}$ for most lines. Similarly, a marginal temporal trend is seen in B268 for all epochs, however, the small number of available double-peaked emission lines makes this detection less significant. The peak-to-peak velocity variations in B275 are relatively small ($<10$\,km\,s$^{-1}$) for all but two \fei/\feii\ lines, which show a similar marginal temporal trend. 

\subsection{(V/R)$_{\rm wing}$-ratio variations}\label{sec:varia_wing}
For B243, the variability detected in the (V/R)$_{\rm wing}$-ratio follows a similar trend for all lines with the exception of \halp\ and \hbet\ (Fig.\,\ref{B243_VR}). However, in the last epoch a decrease in V/R$_{wing}$-ratio is observed for all lines. For the Paschen and Bracket series, the ratio gets as high as 2.2 in the first three epochs and exceeds 3 for O\,{\sc i}, and subsequently decreases towards or below 1, indicating a stronger red side compared to the blue side. The (V/R)$_{\rm wing}$-ratio for \halp\ is below unity for all epochs. 

A detailed of study of the wings of Paschen and Bracket lines in Fig.~\ref{B243_TVS} shows the origin of this temporal trend of the (V/R)$_{\rm wing}$-ratio, especially for the observations in 2019. All Paschen and Bracket wings in the first observation of 2019 (green line, 2019-07-30) show wing excess emission on the blue side compared to the other epochs. Four days later, in the second observation of 2019 (red line, 2019-08-03) this excess in the wing on the blue side has diminished in all Paschen and Bracket lines. Then a few weeks later, during the last observation in 2019 (purple line, 2019-09-25) we observe a wing excess on the red side of the all Paschen and Bracket lines, compared to the other epochs.

We compare this with V/R-ratio variations in B243, that show a similar trend for most lines (Fig.~\ref{B243_VR_peak}). The ratio of the Paschen, Bracket and O\,{\sc i} lines varies between $\sim$0.85 to $\sim$1.3 in a single epoch, which is less than seen for the (V/R)$_{\rm wing}$-ratio. During most of the epochs, most lines have a ratio >~1, indicating a stronger blue than red peak flux. 

In B268 the width of the nebular lines is similar to the peak-to-peak separation in the hydrogen lines. We therefore do not use the V/R-ratio, but focus on the (V/R)$_{\rm wing}$-ratio.
The high velocity (V/R)$_{\rm wing}$-ratio for B268 is >~1 for all lines at all epochs, showing a consistently stronger blue wing than red wing. 
In the case of B275, we do not detect strong temporal variations in the V/R-ratio. However, a strong decrease in the (V/R)$_{\rm wing}$-ratio is observed on the longest observed timescale (2009 to 2019). This is due to a drop in flux on the red side of the features. No significant (V/R)$_{\rm wing}$-ratio variations are observed on timescales of months and days. The exception to this is \hbet, which shows a decrease in ratio on the daily timescale and an increase a few weeks later. 

\section{Discussion}
We present an overview of the characteristics of spectroscopic and photometric variability in our sample of PMS stars and discuss scenarios that could cause this variability. We explore the scenario of a spiral arm in the disk of B243 and B275 based on the systematic differences in spectral lines in a single epoch. The sample is compared with spectroscopic variability in intermediate-mass YSOs to place the PMS stars in M17 in a pre-main-sequence evolutionary context.
The properties of the variability in our sample are used as a diagnostic tool to identify a candidate MYSO in M17.

\subsection{Spectroscopic variability and its possible origins}
Spectroscopic variability comprises:

\begin{enumerate}
    \item[-] differences between the spectra of the studied PMS stars (Fig.~\ref{calcium} and Tab.~\ref{tab:emission_line});
    \item[-] temporal variations between all epochs in three of the four PMS stars with gaseous disks (Fig.~\ref{B243_TVS} - \ref{B275_TVS} and \ref{B337_tvs_Halpha});
    \item[-] differences in the temporal variation of (series of) spectral lines in a single PMS star and among the stars (Fig.~\ref{B243_TVS} - \ref{B275_TVS}).    
\end{enumerate}

The spectral differences and variability hint at the presence of different physical processes ongoing simultaneously in the gaseous circumstellar disks, and suggests that (at least a subset of) these processes may act locally and/or have a time-dependent nature. 

Variations are strongest in double-peaked emission lines and are observed in the peaks and in the wings of these lines. Assuming that the emission originates in a Keplerian rotating disk, the spatial origin of the former is related to where the bulk of the emission is produced and that of the latter to the innermost part of disk. In general, hot gas emitting in the optical and NIR traces the inner regions of PMS stars. Spectral lines of T\,Tauri and Herbig Ae/Be stars in these wavelength domains can display many variable processes due to accretion (bursts), jets, disks structures (that could be remnant of the disk formation process), disk dispersal and/or disk winds, and possibly stellar multiplicity \citep[e.g.,][]{Mora2004, Mendigutia2011, Ellerbroek2014, Scholler2016, contreras-pena2017, Moura2020, Stock2022, Zsidi2022}.

Though a stellar origin of the variability (e.g. starspots, pulsations, flares) cannot be excluded, the high velocities at which the variations occur and the asymmetries in them point to a circumstellar origin.
Therefore, we refrain from further discussion on stellar scenarios and focus instead on possible origins associated with a circumstellar disk.

Mechanisms that we exclude as a possible cause of the observed variability are strong accretion bursts, jets, and close binaries (of similar masses). The first due to a lack of strong photometric variations (we observe a maximum variation of 0.2\,mag, see Sec.~\ref{sec:discus_phot}) and (inverse) P\,Cygni profiles typical for such bursts. Jets are unlikely as we do not detect high-velocity components associated with forbidden emission lines, and for close binarity we lack evidence of radial-velocity variations (>\,$20$\,km\,s$^{-1}$) and/or periodically varying light curves. 

\subsubsection{Typical timescales of spectral variability}
The shortest measured timescale of variability is a few days and it is observed in all stars with spectral variability, in the hydrogen Balmer, Paschen, Bracket series and O\,{\sc i} lines. Daily variability is often stronger than variations between two epochs that are weeks or years apart, though in some targets a pronounced variation on a timescale of years is revealed (e.g. in the Paschen lines of B243 and B275, and the Na\,{\sc i} doublet in B268). Therefore, processes causing the line variability must include mechanisms that manifest on timescales of days and that may be transient in nature, though if so, should have a recurrence time of up to at most years. 

\subsubsection{Relation of peak-to-peak velocity with V/R-ratio}\label{dis:peaktopeak_vs_VR-ratio}
The Paschen and Bracket lines in B243 and Paschen lines in B275 show a positive trend in peak-to-peak velocity with V/R-ratio as displayed in Fig.~\ref{diskvsosc} in a single epoch. 
For B275 this ratio changes in a single epoch from an about equal red and blue peak (V/R $\sim 1$) in the lower series lines \citep[which form over a relatively large region of the disk; see e.g. ][]{Backs2023}, and therefore have lower peak-to-peak velocities) to a stronger blue peak in the higher order H\,{\sc i} lines (which form in the inner regions of the disk, hence have relatively large peak-to-peak velocities).  
This behavior can not be reconciled in the context of an axi-symmetric rotating disk and could imply additional structure in the disk that crosses the line-of-sight at this particular epoch. 

\begin{figure*}
  \centering
  \includegraphics[width=\hsize]{
  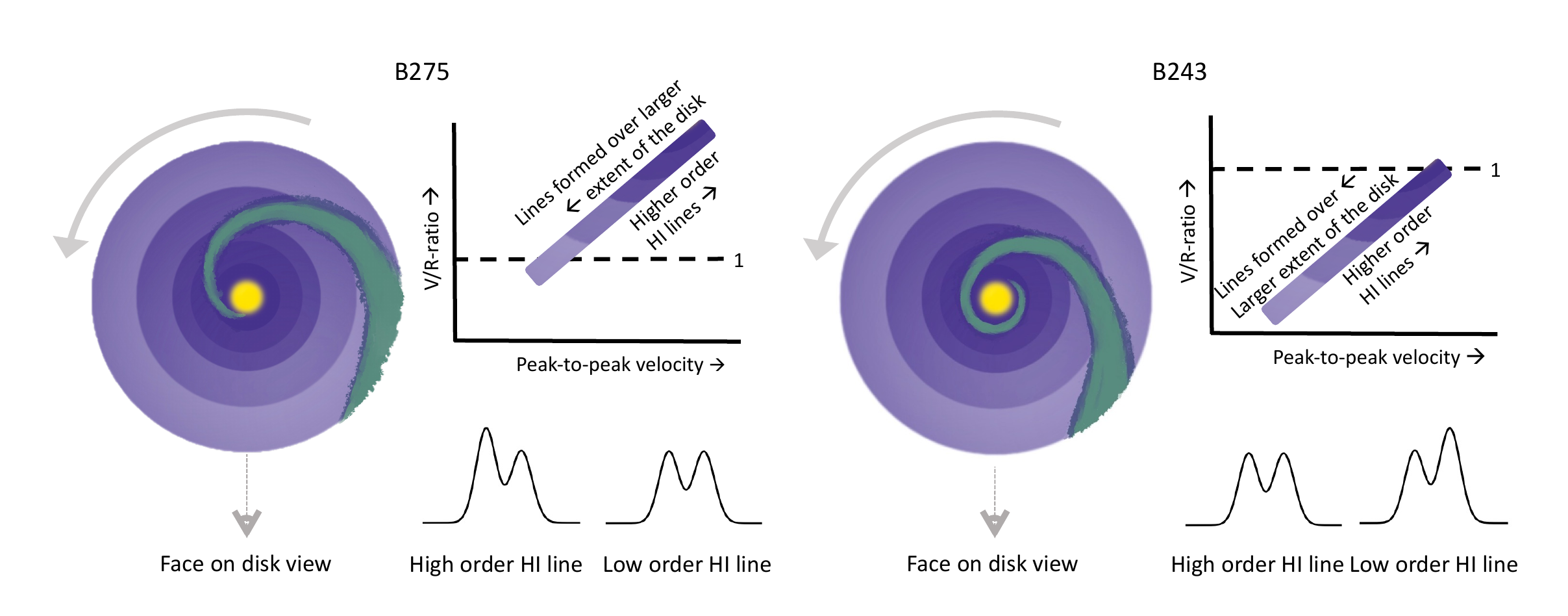}
     \caption{Sketch of the disk for an epoch in B275 and B243. The positive trend between V/R-ratio and peak-to-peak velocity might be explained by a spiral arm in the disks of B275 and B243. We show two examples of phase on inner gaseous disks with a spiral arm and line forming regions in the disk of the higher and lower order H\,{\sc i} lines, where the line region of the latter extend to the surface of the star. Typical double-peaked emission lines are shown for these H\,{\sc i} lines. The figures with V/R-ratio on the y-axis and peak-to-peak velocity on the x-axis mimic the epochs shown in Fig.\,\ref{diskvsosc}. 
             }
        \label{fig:spiralarms}
\end{figure*}

One option that could explain such a trend is a spiral arm in the disk, where a high-density optically-thin or hotter optically-thick structure resides in the more inner parts of the disk (moving toward the observer, hence the stronger blue peak in the high-excitation lines), wavering out to larger distances (explaining the more or less equal red and blue peak strength in the low-excitation lines), see Fig.\,\ref{fig:spiralarms}. For B243 a similar scenario may hold, but with the density enhancement 
of the spiral arm starting on the other side in the inner disk, since the V/R-ratio changes from below unity to about unity and higher in a single epoch. For this star, in 2019 in particular, additional indications exist for a rotating high density structure at high velocities (see \ref{B243_VR}). Nevertheless, the complexity of the line-of-sight towards a PMS star (such as winds, discussed in Sec.~\ref{sec:diskwind}) complicates the identification of such a structure in all epochs, which we do not attempt in this work.

We conclude that for both B275 and B243 there is evidence for disk inhomogeneities 
perhaps similar to V/R-ratio variations tracing the high density part of a one-arm oscillation in Be-star disks \citep{Telting1994}. 
The fact that we detect the V/R-ratio changes in one epoch more clearly as in the others could be an effect of the rotation of the structure with respect to the line-of-sight.
Simulations of (massive) star formation show that the presence of substructures is common in the inner disk regions of massive YSOs and PMS stars that could persist throughout PMS evolution \citep[e.g.,][]{meyer2017, kolligan2018, oliva2020}. We remark, however, that not all line variability characteristics can be explained in this way, leaving room for other phenomena that may be at work simultaneously. 

\subsubsection{Variability at high velocity}\label{dis:diskvel}
\begin{figure}
  \centering
  \includegraphics[width=\hsize]{
  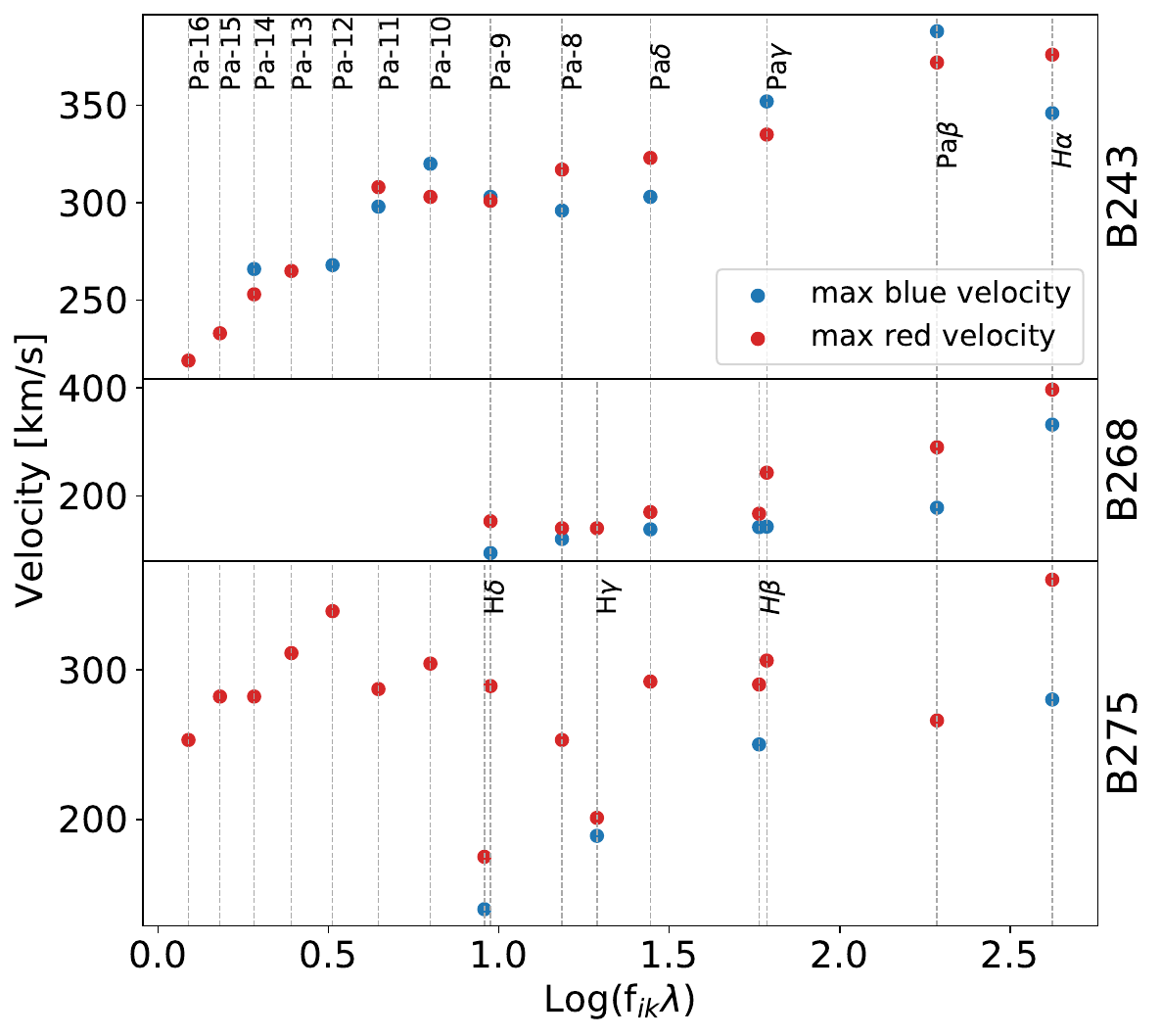}
     \caption{The maximum velocity at which significant variation is observed in the TVS (over all epochs) for each star, plotted against oscillator strength times wavelength. The maximum velocities are determined at the red and at the blue side of the features. We plot the absolute blue side velocity for convenience. The blue velocities of the Paschen lines of B275 are not shown since they do not display variability at high velocities which makes them indistinguishable from nebular subtraction residuals (that range from $\sim$50-80\,km\,s$^{-1}$).
             }
        \label{max_vel}
\end{figure}

To help localize where in the disk the spectroscopic variations in different lines might originate, we measured the maximum velocity at which significant variability is detected in each hydrogen line of the Balmer and Paschen series. This is done both at the blue and at the red side, and plotted against the oscillator strength times wavelength ("line strength") in Fig.~\ref{max_vel}. In B243 and B268, the maximum velocity increases with increasing oscillator strength. This also is the case for the blue side of the Balmer lines in B275, but not for the red side, for which this maximum velocity is fairly constant at about 300\,km\,s$^{-1}$. One could hypothesize that the rising trends are due to an observational bias: for stronger lines it is more easy to detect variability up to higher velocities. However, the trend is not present at the red side in B275.

\begin{figure*}
   \centering
   \makebox[\textwidth]{\includegraphics[width=0.8\textwidth]{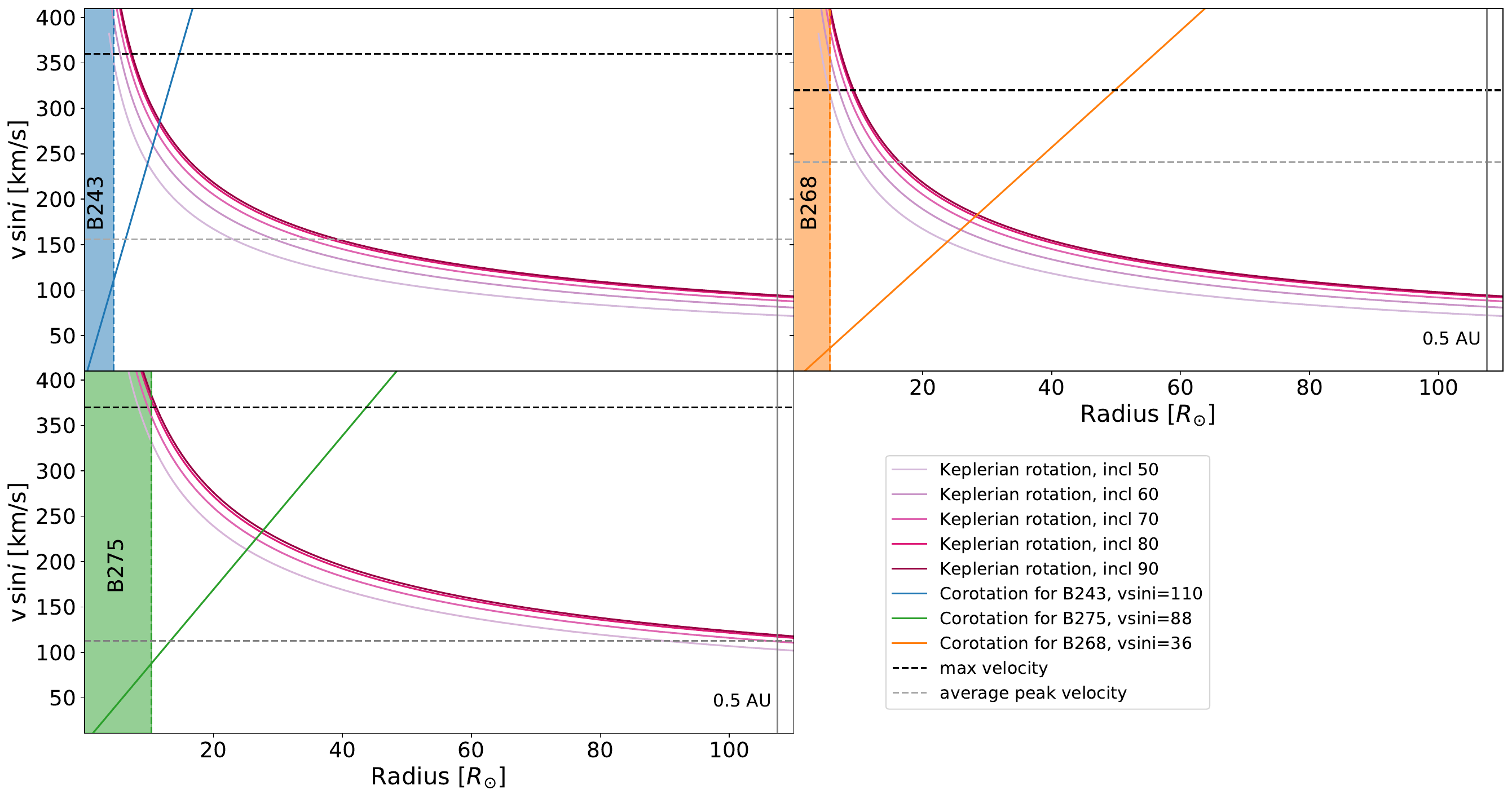}}
      \caption{Projected radial-velocity profiles for various models. The filled-in region on the left of each panel denotes the stellar size. Two velocity-distance relations are plotted; 1. Co-rotation with the stellar surface, where the velocity increases when moving away from the star and 2. Keplerian rotation for several inclinations, where the velocity increase when moving towards the star. The striped lines present the maximum H$\alpha$ velocity detected in the disk (in black), determined by the extend of the variability, and the H$\alpha$ velocity determined by the average of half of the peak-to-peak velocity (in gray).} 
         \label{vels}
   \end{figure*}

In H$\alpha$, variability is detected up to at least 320\,km\,s$^{-1}$ signifying that the inner disk  must rotate with such high velocities. In order to test whether this can be consistent with the lines forming in a Keplerian rotating disk, the velocity expected at a given radius for different inclination angles is shown in Fig.~\ref{vels}. The stellar mass and projected surface rotation velocity $v$\,sin$i$ are taken from Tab.~\ref{tab:MYSO_spec}. In this case, the maximum velocities can be explained with an inclination > 55$^{\circ}$ for B243 and B268, if the disk reaches to the surface of the star. For B275 only an inclination of > 80$^{\circ}$ can explain the highest measured emission line velocities. This is not consistent with the inclination measurement of B275 from \cite{poorta2023} ($\sim$\,60$^{\circ}$) based on modeling of CO bandhead emission. 

Alternatively, one may explain the high velocities by invoking co-rotation (o)f the inner disk), leading to the high velocity emission originating from gas at about 3.5\,$R_{\star}$ in the case of B243, 4\,$R_{\star}$ for B275 and from $\sim$8\,$R_{\star}$ for B268. Co-rotation may be enforced if the disk gas is strongly coupled to a stellar surface magnetic field, as is probably the case in the innermost disk regions of T\,Tauri and Herbig Ae stars, where magneto-spheric accretion may be at play. It is unclear whether magnetic fields play a dominant role in the accretion process of the more massive Herbig Be stars \citep{mendigutia2020, Vioque2022}. In this respect it is noteworthy and surprising that an alternative to a Keplerian velocity profile seems prudent for the most luminous (hence, most massive) source B275. We conclude that for B243 and B268 the high velocity emission may be explained by a disk that extends to close to the surface

As the red side variability is observed up to high velocities in B275, independent of the studied H\,{\sc i} line, and because the maximum H$\alpha$ velocity can only be explained in a Keplerian rotating disk seen edge on, it may imply that the cause of the variability is connected to an alternative origin than from this rotating disk. Possibly the variability at these high velocities is caused by an accretion flow that crosses the line-of-sight. Invoking a magnetically forced co-rotating inner disk up to at least 3.5\,$R_{\star}$ cannot be excluded for explaining the H$\alpha$ velocity. However, this scenario seems less likely, since a visual inspection of the spectral lines in Fig.~\ref{B275_TVS} shows that the spectrum in 2009 displays red-shifted absorption compared to the observations in 2019.

\subsection{Photometric variability}\label{sec:discus_phot}

B243 and B268 do not reveal significant photometric variability ($\Delta\,z^{\prime} \la $\,0.03\,mag; see Fig.~\ref{photometry}). 
For B275, variations $\Delta\,z^{\prime} \sim 0.1-0.2$\,mag are seen on timescales
of hours to days at two of the four observing moments where multiple observations have been secured. T\,Tauri stars, the lower-mass counterpart of our sources, generally show recurring outbursts and photometric variability, often connected to episodes of (increased) mass accretion rates \citep[e.g.,][]{contreras-pena2017, Cody2017, Zsidi2022}, where the amplitude of the photometric variations correlates with the characteristic timescales of the events \citep{Fischer2022}. The strongest outbursts are seen in FU\,Orionis stars (4$-$6\,mag on timescales of years to decades) and EX\,Lupi stars (2$-$4\,mag on timescales of about a year). As we probe similar timescales of years to a decade, these types of outburst are likely excluded for our sources. For T\,Tauri stars, 0.5$-$2\,mag outbursts have been observed during magnetospheric accretion events. This type of irregular and smaller accretion events could be responsible for the small shifts in 
$\Delta\,z^{\prime}$ in B275, as it has been shown that the amplitude of the photometric variations may depend on the band that is probed \citep{Ellerbroek2014} as well as on the system's inclination \citep{Kesseli2016}. 
However, it may also be related to other phenomena that operate on hours/days timescales, such as stellar flares, pulsations, small companions or starspots if the star is magnetic, or other phenomena that operate on dynamical or rotational timescales of the star or the very inner disk \citep{Fischer2022}.

\subsection{Potential origin of the emission lines}\label{sec:source_varia_diff}
Though the line variability behavior is complex, certain lines show variations that appear correlated in time as well as in strength. Specifically, these are the O\,{\sc i} line at 844.6\,nm and 1128.7\,nm, C\,{\sc i} and H\,{\sc i} emission in B243, B275, and (weakly) in B337, and the Ca\,{\sc ii} triplet, Mg\,{\sc i}, and Fe\,{\sc i} at 999.8\,nm emission in B268, B275, and B337 (see Tab.~\ref{tab:emission_line}). 
Likely, these correlations could imply that the same mechanism is responsible for the variations.

The simultaneous presence of strong O\,{\sc i} and H\,{\sc i} lines is due to Bowen resonance fluorescence, where Ly$\beta$ photons excite the O\,{\sc i} resonance line 2$p$ $^{3}P_2-3d$ $^{3}D^{0}_{321}$ at 102.58\,nm, from which they cascade through the O\,{\sc i} at 844.6\,nm and 1128.7\,nm lines follow. This has been identified as the dominant excitation mechanism of O\,{\sc i} lines in Herbig Be stars \citep{Mathew2018} and is a well-known phenomenon in H\,{\sc ii} regions. The link implies that O\,{\sc i} emission is co-spatial with (optically thick) H\,{\sc i} Lyman excitation, where H$\alpha$ is used by \cite{Mathew2018} as a tracer for Lyman excitation.

For B243 and B275, the velocity regime in which variability is detected for H$\alpha$ and O\,{\sc i} agree (Fig.~\ref{B243_TVS} and \ref{B275_TVS}), further underpinning this connection. Both stars also show double-peaked C\,{\sc i} emission. To our knowledge, there has not been an association of C\,{\sc i} emission with any pumping by hydrogen or helium resonance lines, similar to O\,{\sc i}. We note that the stars with the strongest O\,{\sc i} emission also show a handful of double-peaked C\,{\sc i} emission lines. This may imply a relation, yet to be investigated, with the (destruction of) CO molecules, that are observed as CO-bandhead emission in both stars. So far in literature, C\,{\sc i} observations are only discussed in the context of T\,Tauri stars where it is used as a parameter to characterize the formation of planets \citep{mcclure2019}. 

We do not observe variations in the CO-bandhead emission, which consists of super positions of many CO transitions, complicating the detection of variability. However, relatively large changes in the disk; in inclination, mid plane disk density or mid plane disk temperature where the CO bandheads are thought to form \citep{ilee2018, poorta2023}, should be observable as e.g. seen in an extreme case during an accretion outburst \citep{caratti2017}. 

A strong Ca\,{\sc ii} triplet and Mg\,{\sc i} and Fe\,{\sc i} at 999.8\,nm emission are simultaneously present in B268, B275 and B337, but none of the lines appear (strongly) in B243. The variability in the Ca\,{\sc ii} triplet emission is characterized by variability in the peaks, both in B268 and B275. The other lines do not show variations. 

The variability in the Ca\,{\sc ii} triplet and O\,{\sc i} lines are not correlated, which is similar to what is observed in Be stars \citep{Shokry2018}. These authors conclude that the origin of the variability in the two species should be attributed to processes that are spatially separated, possibly caused by independent mechanisms acting in the disk.

B268 is the only star with superimposed Na\,{\sc i} (D1 and D2) absorption lines on the sharp Na\,{\sc i} absorption lines that arise from the ISM. Variability is detected in these lines between 2012/2013 and 2019 where the absorption has disappeared. Variability in the Na\,{\sc i} lines is also observed in intermediate mass YSOs \citep{Mora2004, Mendigutia2011}, where the variations are linked to the circumstellar disk.

\subsection{Considerations regarding mass-accretion rate estimates} \label{sec:discus_massacc}
As part of the final stage of formation, PMS stars undergo accretion. The rate of accretion $\dot{M}_{\rm acc}$ may help pinpoint the evolutionary phase of the system and, when in-fall of material from the natal cloud has ceased, the remaining lifetime of the disk. Mass-accretion rates of T\,Tauri and Herbig stars are estimated from ultraviolet (UV) or Balmer-continuum excess, thought to be due to gas, funneled along magnetic field lines that connect the inner disk to the star, that shocks the photosphere upon impact. The exact relation between $\dot{M}_{\rm acc}$ and the additional luminosity caused by accretion $L_{\rm acc}$ depends on several factors, including the velocity of impact, the fraction of surface filled by accretion columns, the quiet surface stellar temperature, and the inner disk radius \citep[e.g.,][]{fairlamb2015}. This additional luminosity caused by accretion is empirically found to correlate with emission line luminosity \citep[e.g.,][]{Mendigutia2011,fairlamb2017}. This does not suggest that all the line emission originates from the accretion stream; it may originate from both the in-falling gas and gas that is in Keplerian orbit in the inner disk. For instance, \citet{patel2017} and \citet{Backs2023} show that hydrogen emission lines can be reproduced by adopting a stationary Keplerian disk model only. 
The scatter in $L_{\rm acc}-L_{\rm line}$ relations typically amounts to an order of magnitude \citep[e.g.,][]{fairlamb2017, Vioque2022}. This implies that care should be taken when interpreting the resulting mass-accretion rates, especially when no direct indications of mass accretion are observed.

\begin{figure}
   \centering
   \includegraphics[width=\hsize]{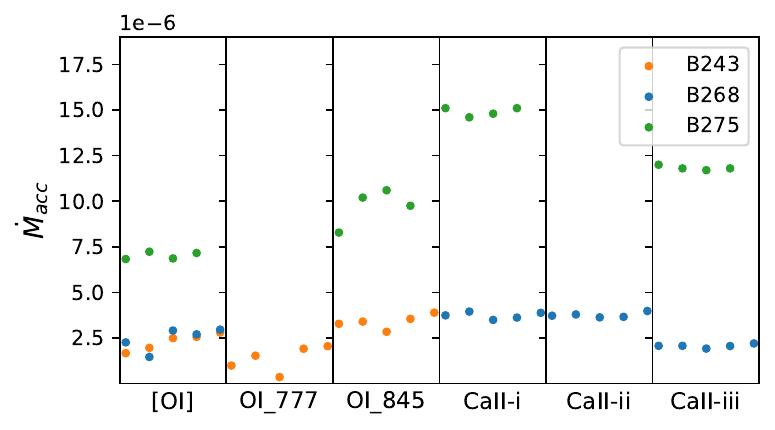}
      \caption{Accretion rates from spectral line luminosities in M$_{\odot}$\,yr$^{-1}$ for B243, B268 and B275. Each bracketed column is for a given line and shows the rates for all epochs separately. The values should be considered with care given the uncertainties associated with the applied method (see text), but typically are of the order of $\sim 10^{-6}$ M$_{\odot}$ yr$^{-1}$ for B243 and B268, and up to $\sim 10^{-5}$ M$_{\odot}$ yr$^{-1}$ for B275.}
         \label{Macc}
\end{figure}

For our sources UV or Balmer continuum photometry is not available, therefore we resort to using line luminosities to obtain mass-accretion rates. We use the lines that have an equivalent width measurement in Tab.~\ref{tab:EW} and SEDs from \RT\ to obtain the absolute line luminosities, and apply the relations provided in \cite{fairlamb2017}. The resulting accretion rates for each epoch are presented in Fig.~\ref{Macc} and Tab.~\ref{tab:mass_acc}. We obtain typical values of $\sim$10$^{-6}$-10$^{-5}$ M$_{\odot}$ yr$^{-1}$, comparable to what is found for Herbig Be stars \citep{Mendigutia2011, Mendigutia2013, fairlamb2015, ababakr2017, wichittanakom2020, Vioque2022, Brittain2023}.

Adopting a typical time for our sources to arrive on the ZAMS of $\sim$10$^{5}$\,yr (see Fig.~\ref{HRD}), the disk should have a mass of $\sim$0.1\,$M_{\odot}$ to remain present for the full span of this remaining PMS evolution. Recent estimates of the masses of the disks of B243, B268 and B275 from ALMA continuum observations yield masses of order $10^{-3}\,M_{\odot}$ only \citep{poorta2:inpress-a}, implying that these disks will be cleared in $\sim$\,100-1000\,yr for these derived mass-accretion rates. It seems unlikely that we picked up these stars at such a critical time. This further emphasizes that the derived $\dot{M}_{\rm acc}$ must be handled with care.

\subsection{Slow disk winds and jets}\label{sec:diskwind}

In T\,Tauri stars, the shape and strength of (blue-shifted) observed forbidden lines \citep{Natta2014}, notably [O\,{\sc i}] at 630\,nm, probably point to the escape of gas from the PMS system, either in a slow wind emanating from the disk surface \citep{Nisini2018} or in high-velocity jets piercing in the polar directions. B243 and B275, and tentatively B268, have [O\,{\sc i}]\,at 630\,nm in emission (see Fig.~\ref{missing_lines}). Velocities where the line is in emission range from about [-50, 50]\,km\,s$^{-1}$. When the peak is observed at these velocities, close to the system velocity, as is the case here, it is for T\,Tauri stars classified as a low-velocity component \citep[][]{1987ApJ...321..473E,Nisini2018}. Systems that show emission lines with a high-velocity component (HVS; $> 200$\,km\,s$^{-1}$), likely feature jets that arise from the mass-accretion process \citep{Ellerbroek2014}.

An origin of [O\,{\sc i}] at 630\,nm in a flow that is launched away from the disk surface is supported by the single-peaked nature of the line (while allowed lines coming from the disk are typically double peaked). For a disk inclination of 70$^{\circ}$, a jet with velocities up to 146\,km\,s$^{-1}$ beaming in the direction normal to the disk surface would also show a projected velocity of 50\,km\,s$^{-1}$.
However, jets are usually associated with early stages of PMS evolution and seem at odds with the apparently low disk masses inferred by \citet{poorta2:inpress-a}. Often these jets give rise to multiple forbidden-line transitions, like [F\,{\sc ii}], [Ni\,{\sc ii}], [S\,{\sc ii}] \citep{Ellerbroek2014}, which are not observed in the PMS stars. 
A slow disk wind in B243, B268 and B275 seems a more plausible outflow model assuming that T\,Tauri stars and more massive PMS stars have similar disk wind/jet mechanisms.

\subsection{Variability as a mean to identify circumstellar material}
From all six PMS stars, three of the four stars with gaseous disks show spectroscopic variations in their optical and NIR emission lines at relatively high velocities (>\,100\,km\,s$^{-1}$). There is no observed variability in the spectra of B215 and B289 (Fig.~\ref{B215_tvs}-\ref{B289_tvs}), that do not show emission lines (outside of the range affected by nebular lines >\,50\,km\,s$^{-1}$). B337 has emission lines, but also shows strong nebular subtraction residuals and is heavily extincted, restricting the detection of variability in this star (Fig.~\ref{B337_tvs_Halpha}). This result shows that variability from the hot disk gas could be a common property in PMS stars with gaseous disks in M17 and also that strong nebular emission in M17 complicates the detection of emission for stars with emission lines with velocities $\sim$50-100\,km\,s$^{-1}$ (e.g. that originate from a fairly face-on disk). Assuming that this type of spectroscopic variability is a more common property in PMS stars, it could be used as a diagnostic to identify stars with gaseous circumstellar material.

\begin{figure}
  \centering
  \includegraphics[width=0.8\hsize]{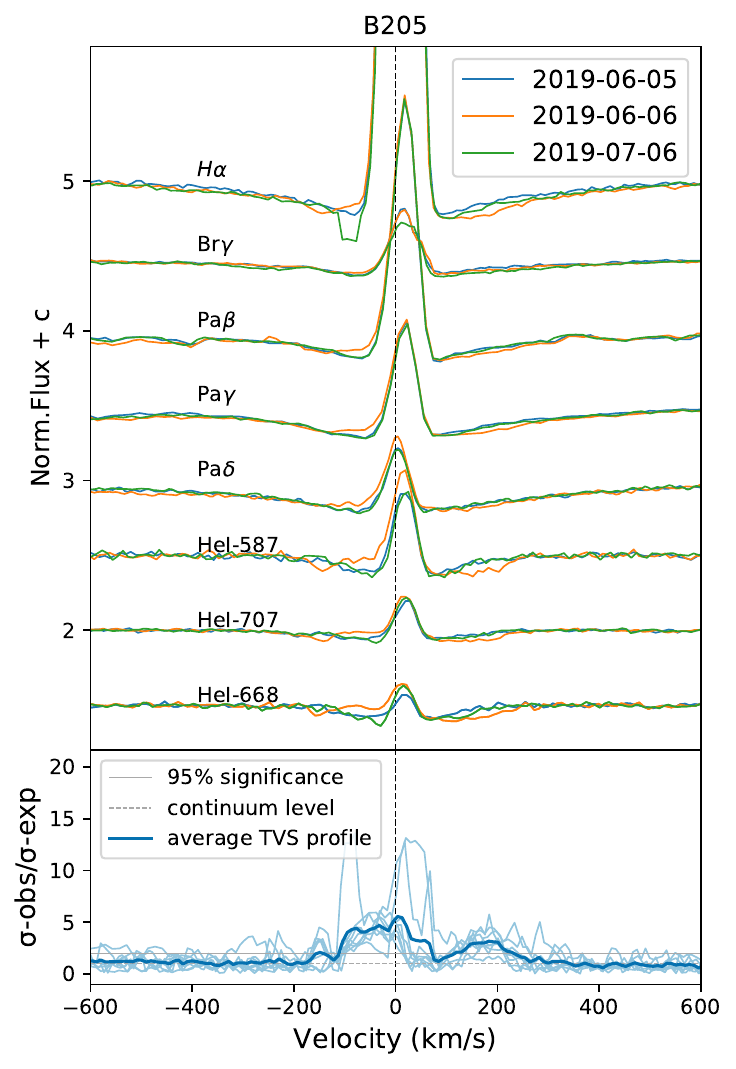}
     \caption{Lines profiles and TVS of B205. The upper panel shows Balmer, Paschen, Bracket and Helium lines in several spectra of B205 taken in 2019. The lower panel shows the TVS of these spectra, indicating a significant velocity range between -150\,km\,s$^{-1}$ and 220\,km\,s$^{-1}$. Variations in the spectra look strongest in the helium lines. The emission around the rest wavelength of the features is of nebular origin confirmed by their 2D spectra. 
             }
        \label{tvs_B205}
\end{figure}

To test this potential diagnostic, we apply it to identify circumstellar material around the other stars from the spectroscopic monitoring campaign of OB stars in M17 (see Sec.~\ref{sec:sample}). Using the TVS method, one of the stars, B205, shows significant spectroscopic variations in multiple lines up to 200\,km\,s$^{-1}$ (Fig.~\ref{tvs_B205}). This star has been classified by \cite{Povich2017} as a B2 star by its NIR colors and J-band magnitude. Though \cite{Hanson1997} did not detect infrared excess for this source, \cite{backs:inpress-a} place this star on the PMS track in the HRD. It could be that this star still features a remnant gaseous disk, challenging our understanding of disk removal, or this star could have developed, despite the young age of M17 ($\sim$0.65$\pm$0.25\,Myr \citep{stoop_inprep}), into a Be star. More investigation is needed to confirm or reject a pre-main-sequence nature for B205.  

\section{Conclusion}
This paper presents a multi-epoch study of simultaneous spectroscopic and photometric observations of six PMS stars in M17; four with gaseous disks and two PMS stars with IR excess redwards of 3.0\,$\mu$m. These PMS stars with observable photospheres and gaseous disks exhibit intrinsic variability (except the most embedded one). 

Spectroscopic variability is observed on all timescales (especially in strong emission lines) that range from days to years and seem transient over the longer timescales of months and years. The extent and timescale of the variability differs per line and object which shows the complexity of the physical conditions in the regions where the different lines are formed. This leads us to believe that multiple processes are at play, causing variations in these gaseous disks. 

In one epoch of B243 and B275 we observe a positive trend of the V/R-ratio with projected rotational velocity as measured from different double-peaked emission lines. In both cases V/R-ratio takes values above and below unity, which in this particular case might hint on a structure in the disk that crosses the line-of-sight; like a spiral arm. Other differences in spectral line set show that two similar PMS stars (B243 and B268) with the same central object in terms of spectral type, age and mass show differences in disk properties. 

The weak peak-to-peak velocity variations between the epochs and the consistency in the double peak morphology for lines in each epoch show that the bulk of the probed disk region is relatively stable or that temporal effects are adequately smeared out. This is supported by the fact that we do not observe variations in the CO bandhead emission. Variability is largely confined to the disk region close to the surface of the star and disk structures on top of or in the relatively stable bulk of the disk.

Determining the maximum velocity of variation shows that variability is detected at high velocities (>\,320\,km\,s$^{-1}$), indicating an origin close to the surface of the stars. For B275, the high variable velocities on the red side of the H\,{\sc i} lines, independent of H\,{\sc i} line strength, hint at an origin of the variability in an accretion flow from the inner disk region to the star. However, the lack of strong photometric variations (in all stars) shows that this flow cannot cover a significant fraction of the star/disk and large accretion bursts (EX Lupi or FU Orionis type) are excluded.  

The presence of a low velocity component in [O\,{\sc i}] at 630\,nm that does not vary, shows that we have to take in account relatively slow outflows/winds that could be launched from the disk. 

Our PMS stars do not seem to fit the empirically determined $\dot{M}_{\rm acc} - L_{\rm acc}$ relation. When assuming that these objects are still accreting, estimated mass accretion rates 
show much higher accretion rates than expected for these PMS stars. This might be due to a different accretion mechanism for the more massive Herbig stars, accompanied by a different $\dot{M}_{\rm acc} - L_{\rm acc}$ relation, presently not accounted for in mass-accretion estimates for the more massive PMS stars.

The spectroscopic variability shows there are multiple processes at play in these PMS stars with gaseous disks. The next step is to obtain more multi-epoch spectroscopy of a larger sample of stars to tie variability to a certain phase in the pre-main-sequence evolution and to more specifically link variations to certain disk processes. 

\begin{acknowledgements}
We thank the anonymous referee for carefully reading the manuscript and many helpful and insightful comments and suggestions. We acknowledge support from the Netherlands Research School for Astronomy (NOVA). This research has made use of the SIMBAD database, operated at CDS, Strasbourg, France \citep{wenger2000}. This research made use of Astropy, a community-developed core Python package for Astronomy \citep{astropy2022}. This work has made use of data from the European Space Agency (ESA) mission Gaia (https://www. cosmos.esa.int/gaia), processed by the Gaia Data Processing and Analysis Consortium (DPAC, https://www.cosmos.esa.int/web/gaia/dpac/ consortium). Funding for the DPAC has been provided by national institutions, in particular the institutions participating in the Gaia Multilateral Agreement.
\end{acknowledgements}

%
%
\bibliographystyle{aa}
\bibliography{main}

\begin{appendix}

\section{Observations}

The observations for each star and date are listed in the table below and show exposure times for each observation in a particular arm. 

\begin{table}[]
\small
\caption{Exposure times and S/N for each observation.}
\label{obslog}
\begin{tabular}{lllrl} 
\toprule 
\midrule
 Star &       Date &  Arm &   Exp.time (s) & S/N \\
\midrule
\midrule

 B243 &   2012-07-06 &  UVB &  4x900 & 15 \\
 &    &  VIS &  4x870 & 152\\
 &    &  NIR &   4x50 & 366 \\
  \midrule
 &   2013-07-17 &  UVB &  2x870  & 19\\
 &    &  VIS &  4x450 & 149 \\
 &    &  NIR &   2x50 & 290 \\
  \midrule
 &   2019-07-30 &  UVB &  4x708  & 29 \\
 &    &  VIS &  4x737 & 145 \\
 &    &  NIR &   6x50 & 260 \\
  \midrule
 &   2019-08-03 &  UVB &  4x708  & 19\\
 &    &  VIS &  4x737 & 149 \\
 &    &  NIR &   6x50 & 231\\
  \midrule
 &   2019-09-25 &  UVB &  4x708 & 26\\
 &    &  VIS &  4x737 & 169 \\
 &    &  NIR &   6x50 & 236 \\
 \midrule
 \midrule
 B268 &   2012-07-06 &  UVB &  4x870 & 35 \\
  &      &  VIS &  8x300 & 235 \\
  &      &  NIR &   4x50 & 162 \\
   \midrule
  &     2013-07-17 &  UVB &  2x870  & 43 \\
  &      &  VIS &  2x450 & 168 \\
  &      &  NIR &   2x50 & 275 \\
   \midrule
  &     2019-05-30 &  UVB &  4x708 & 47\\
  &      &  VIS &  4x737 & 172 \\
  &      &  NIR &   6x50 & 268 \\
   \midrule
  &     2019-05-31 &  UVB &  4x708 & 51 \\
  &      &  VIS &  4x737 & 246 \\
  &      &  NIR &   6x50 & 308 \\
   \midrule
  &     2019-07-29 &  UVB &  4x708 & 34 \\
  &      &  VIS &  4x737 & 120 \\
  &      &  NIR &   6x50 & 119 \\
  \midrule
   \midrule
 B275  &  2009-08-11 &  UVB &  4x685 & 132\\
  &      &  VIS &  8x285 & 212\\
  &      &  NIR &  12x11 & 45\\
   \midrule
  &     2019-06-05 &  UVB &  4x350 & 68\\
  &      &  VIS &  4x300 & 252\\
  &      &  NIR &  18x10 & 178 \\
   \midrule
  &     2019-06-06 &  UVB &  4x350 & 92 \\
  &      &  VIS &  4x300 & 362\\
  &      &  NIR &  18x10 & 150 \\
   \midrule
  &     2019-07-09 &  UVB &  4x350 & 93\\
  &     &  VIS &  4x300 & 351\\
  &     &  NIR &  18x10 & 149 \\
  \midrule
   \midrule
 B337 &  2013-07-16 &  UVB &  4x870 & 0.4\\
  &    &  VIS &  8x450 & 73 \\
  &    &  NIR &   4x50 & 177 \\
   \midrule
  &   2019-05-01 &  UVB &  4x708  & 0.02  \\
  &    &  VIS &  4x737 & 27 \\
  &    &  NIR &   6x50 & 231 \\
   \midrule
  &   2019-08-03 &  UVB &  4x708 & 0.1 \\
  &    &  VIS &  4x737 & 35\\
  &    &  NIR &   6x50 & 260 \\
  \bottomrule
\end{tabular}
\begin{tablenotes}
    \small
    \item \textbf{Notes.}  The S/N is calculated in the continuum of the UVB arm at $\sim468$\,nm, in the VIS arm at $\sim808$\,nm and the NIR arm at $\sim1537$\,nm.
\end{tablenotes}
\end{table}

\section{Circumstellar lines}\label{app:spectral_lines}
We provide a detailed discussion of emission and absorption
lines in the spectrum for each of the four PMS stars with a
gaseous disk.

\twocolumn
\begin{sidewaystable*}[h]
\caption{Lines detected in the spectra (star + circumstellar) of B243, B268, B275 and B337. }\label{YSOtable}
\centering
\begin{tabular}{@{}
l 
l |
l 
l 
l 
l ||
l 
l |
l 
l 
l 
l ||
l 
l |
l 
l 
l 
l @{}}
\toprule \toprule
\textbf{} &
  \textbf{$\lambda$ (nm)} &
  \textbf{B243} &
  \textbf{B268} &
  \textbf{B275} &
  \textbf{B337} &
   &
  \textbf{$\lambda$ (nm)} &
  \textbf{B243} &
  \textbf{B268} &
  \textbf{B275} &
  \textbf{B337} &
   &
  \textbf{$\lambda$ (nm)} &
  \textbf{B243} &
  \textbf{B268} &
  \textbf{B275} &
  \textbf{B337} \\ \midrule \midrule
\textbf{H$\eta$} &
  383.5397 &
  A &
  A &
  A &
  - &
  \textbf{} &
  666.0 &
  A &
  A &
  A &
  - &
  \textbf{C\,{\sc   i}} &
  965.84 &
  E-d &
   &
  E-d &
  - \\
\textbf{Ca\,{\sc ii}} &
  393.4 &
  A &
  A$^{\rm V}$ &
  A &
  - &
  \textbf{} &
  743.0 &
  A &
  A &
  A &
  A &
  \textbf{Fe\,{\sc i}} &
  999.8 &
  - &
  E-s &
  E-s &
  E-s \\
\textbf{Ba-7} &
  397.0075 &
  A &
  A &
  A &
  - &
  \textbf{Fe\,{\sc ii}} &
  746.3 &
  - &
  A &
  E-d &
  - &
  \textbf{Pa$\delta$} &
  1004.98 &
  B$^{\rm V}$-d &
  B$^{\rm V}$ &
  B$^{\rm V}$ &
  B \\
\textbf{H$\delta$} &
  410.1734 &
  A &
  A &
  B$^{\rm V}$ &
  - &
  \textbf{Fe\,{\sc ii}} &
  751.4 &
  - &
  - &
  E-d &
  - &
  \textbf{C\,{\sc i}} &
  1012.38 &
  E-d &
  - &
  E-d &
  - \\
\textbf{CH+} &
  423.3 &
  - &
  - &
  A &
  - &
  \textbf{Fe\,{\sc ii}} &
  771.2 &
  - &
  A &
  E-d &
  - &
  \textbf{Fe\,{\sc i}} &
  1045.8 &
  - &
  - &
  E-d &
  - \\
\textbf{H$\gamma$} &
  434.0472 &
  B &
  A &
  B$^{\rm V}$ &
  - &
  \textbf{O\,{\sc i}} &
  777.3387 &
  E$^{\rm V}$ &
  A$^{\rm V}$ &
  A$^{\rm V}$ &
  A &
  \textbf{Fe\,{\sc ii}} &
  1050.1 &
  - &
  - &
  E-d &
  - \\
\textbf{Mg\,{\sc ii}} &
  448.11 &
  A &
  A &
  A &
  - &
  \textbf{O\,{\sc i}} &
  844.636 &
  E$^{\rm V}$-d &
  - &
  E$^{\rm V}$-d &
  E-d &
  \textbf{C\,{\sc i}} &
  1068.83 &
  E-d &
  - &
  E-d &
  E-d \\
\textbf{} &
  455.0 &
  - &
  A &
  - &
  - &
  \textbf{Ca\,{\sc ii}} &
  849.802 &
  E &
  E$^{\rm V}$-d &
  E$^{\rm V}$-d &
  E-d &
  \textbf{C\,{\sc i}} &
  1072.95 &
  E &
  - &
  E &
  - \\
\textbf{} &
  458.8 &
  - &
  A &
  - &
  - &
  \textbf{Pa-16} &
  850.249 &
  E$^{\rm V}$-d &
  A$^{\rm V}$ &
  A &
  A &
  \textbf{He\,{\sc i}} &
  1086.8 &
  - &
  - &
  E &
  - \\
\textbf{H$\beta$} &
  486.135 &
  B$^{\rm V}$ &
  A &
  B$^{\rm V}$ &
  - &
  \textbf{Ca\,{\sc ii}} &
  854.209 &
  E &
  E$^{\rm V}$-d &
  E$^{\rm V}$-d &
  E-d &
  \textbf{Pa$\gamma$} &
  1093.817 &
  B$^{\rm V}$-d &
  B$^{\rm V}$ &
  B$^{\rm V}$ &
  B \\
\textbf{Fe\,{\sc ii}} &
  501.5 &
  A &
  A$^{\rm V}$ &
  A &
  - &
  \textbf{Pa-15} &
  854.538 &
  E$^{\rm V}$-d &
  A$^{\rm V}$ &
  A &
  A &
  \textbf{O\,{\sc i}} &
  1128.7 &
  E-d &
  - &
  E-d &
  - \\
\textbf{Fe\,{\sc i}} &
  505.60 &
  - &
  A &
  A &
  - &
  \textbf{Pa-14} &
  859.839 &
  B$^{\rm V}$-d &
  A &
  B$^{\rm V}$-d &
  A &
  \textbf{C\,{\sc i}} &
  1175.33 &
  E-d &
  - &
  E-d &
  - \\
\textbf{Fe\,{\sc i}} &
  516.98 &
  - &
  A$^{\rm V}$ &
  E$^{\rm V}$-d &
  - &
  \textbf{Ca\,{\sc ii}} &
  866.214 &
  E &
  E$^{\rm V}$-d &
  E$^{\rm V}$ &
  E-d &
  \textbf{C\,{\sc i}} &
  1189.58 &
  E-d &
  - &
  E-d &
  - \\
\textbf{Fe\,{\sc ii}} &
  519.6 &
  - &
  A$^{\rm V}$ &
  E-d &
  - &
  \textbf{Pa-13} &
  866.502 &
  E$^{\rm V}$-d &
  A$^{\rm V}$ &
  A$^{\rm V}$ &
  A &
  \textbf{} &
  1227.10 &
  - &
  - &
  E &
  - \\
\textbf{} &
  523.5 &
  - &
  A &
  - &
  - &
  \textbf{Fe\,{\sc i}} &
  868.8 &
  A$^{\rm V}$ &
  A &
  E-d &
  - &
  \textbf{} &
  1233.80 &
  A &
  A &
  A &
  A \\
\textbf{} &
  527.6 &
  - &
  A &
  - &
  - &
  \textbf{Pa-12} &
  875.0 &
  B$^{\rm V}$-d &
  B &
  B$^{\rm V}$-d &
  B &
  \textbf{Pa$\beta$} &
  1281.8072 &
  B$^{\rm V}$-d &
  B$^{\rm V}$ &
  B$^{\rm V}$ &
  B \\
\textbf{Fe\,{\sc ii}} &
  531.7 &
  - &
  A$^{\rm V}$ &
  E-d &
  - &
  \textbf{Mg\,{\sc i}} &
  880.8 &
  - &
  E-d &
  E-d &
  E-d &
  \textbf{C\,{\sc i}} &
  1454.24 &
  E-d &
  - &
  E-d &
  - \\
\textbf{} &
  570.6 &
  A &
  A &
  A &
  - &
  \textbf{Fe\,{\sc i}} &
  882.4 &
  - &
  E-d &
  E-s &
  - &
  \textbf{Br-16} &
  1555.621 &
  B-d &
  A &
  B &
  A \\
\textbf{He\,{\sc i}} &
  587.64 &
  B$^{\rm V}$ &
  - &
  B &
  - &
  \textbf{Pa-11} &
  886.3 &
  B$^{\rm V}$-d &
  B &
  B$^{\rm V}$-d &
  A &
  \textbf{Br-14} &
  1588.0558 &
  B$^{\rm V}$-d &
  A &
  B &
  A \\
\textbf{Na\,{\sc i}} &
  589.64 &
  A &
  A$^{\rm V}$ &
  A &
  - &
  \textbf{Pa-10} &
  901.5 &
  B$^{\rm V}$-d &
  B &
  B-d &
  B &
  \textbf{Br-12} &
  1640.688 &
  B$^{\rm V}$ &
  A &
  B$^{\rm V}$-d &
  A \\
\textbf{} &
  601.1 &
  A &
  A &
  A &
  - &
  \textbf{C\,{\sc i}} &
  906.19 &
  E$^{\rm V}$-d &
  - &
  E$^{\rm V}$-d &
  - &
  \textbf{Br-11} &
  1680.651 &
  B$^{\rm V}$-d &
  B &
  B &
  B \\
\textbf{Fe\,{\sc i}} &
  628.4 &
  A &
  A &
  A &
  - &
  \textbf{C\,{\sc i}} &
  909.48 &
  E-d &
  - &
  E-d &
  - &
  \textbf{Fe\,{\sc i}} &
  1687.3 &
  - &
  E-s &
  E-s &
  E-s \\
\textbf{[O\,{\sc i}]} &
  630.0304 &
  E-s &
  E-s &
  E-s &
  - &
  \textbf{C\,{\sc i}} &
  911.18 &
  E-d &
  - &
  E-d &
  - &
  \textbf{C\,{\sc i}} &
  1689.03 &
  E-d &
  - &
  E-d &
  - \\
\textbf{Fe\,{\sc ii}} &
  634.7 &
  A &
  A &
  A &
  - &
  \textbf{Pa-9} &
  922.9 &
  B$^{\rm V}$-d &
  B &
  B$^{\rm V}$-d &
  B &
  \textbf{Br-10} &
  1736.2143 &
  B$^{\rm V}$-d &
  B &
  B &
  B \\
\textbf{} &
  645.6 &
  A &
  A &
  - &
  - &
  \textbf{O\,{\sc i}} &
  926.4 &
  A$^{\rm V}$ &
  A &
  A &
  A &
  \textbf{Br$\delta$} &
  1944.5581 &
  B-d &
  B &
  B &
  B \\
\textbf{Fe\,{\sc ii}} &
  651.6 &
  - &
  E-d &
  E-d &
  - &
  \textbf{C\,{\sc i}} &
  940.57 &
  E-d &
  - &
  E-d &
  - &
  \textbf{Br$\gamma$} &
  2165.5268 &
  B$^{\rm V}$-d &
  B$^{\rm V}$ &
  B$^{\rm V}$ &
  B \\
\textbf{H$\alpha$} &
  656.279 &
  B$^{\rm V}$ &
  B$^{\rm V}$ &
  B$^{\rm V}$ &
  B &
  \textbf{Pa-8} &
  954.62 &
  B$^{\rm V}$-d &
  B &
  B$^{\rm V}$-d &
  A &
   &
   &
   &
   &
   &
   \\ \bottomrule
\end{tabular}
\begin{tablenotes}
    \small
    \item \textbf{Notes.} A indicates the detection of an absorption line, E of emission and B of both absorption and emission. The superscript V specifies that the line is varying over the epochs. An additional s or d is used when the emission in the line is clearly double- or single-peaked. Some lines are detected but not identified.
\end{tablenotes}
\end{sidewaystable*}

\twocolumn
\subsection{B243}\label{p1:app:B243}
The spectrum of B243 is characterized by (double) peaked line emission in H\,{\sc i}, Ca\,{\sc ii} and O\,{\sc i}. The Paschen and Bracket absorption lines are superimposed with two clearly separated emission peaks. Only the strongest Balmer lines are partly visible in emission. 

In particular, while photospheric wings are detectable as blue as Ba-13/H$\theta$, emission is not observed in H$\gamma$ and the bluer H\,{\sc i} lines. A narrow residual from the nebular subtraction is still present in all H\,{\sc i} lines. For the Paschen series this is visible as a small peak of ${\sim}50$\,km\,s$^{-1}$ in width around the rest wavelength, while for the Balmer series the contamination is broader with a width of ${\sim}160$\,km\,s$^{-1}$ at the bottom of the feature. This relatively broad range compared to the intrinsic width of nebular lines is caused by variations of the nebular emission lines along the slit, increasing the affected region during the subtraction of these lines, see Sec.~\ref{sec:neb}.

The double-peaked Ca\,{\sc ii} triplet is weak compared to the Paschen lines. The features blend, therefore, only a weak blue peak of the Ca\,{\sc ii} appears in the wings of the Paschen lines. O\,{\sc i} emission is observed at 844.6\,nm, double-peaked in emission, likewise for the triplet at 777.4\,nm, but less strong. Additionally, the O\,{\sc i} triplet is present in absorption at 926.2-926.6\,nm. Lastly, forbidden [O\,{\sc i}] at 630.0\,nm appears single-peaked in all of the spectra in emission. The C\,{\sc i} lines appear double-peaked in emission for wavelengths longer than 900\,nm. B243 shows weak CO bandhead emission compared to B268 and B275. 

\subsection{B268}
The inner core of the H\,{\sc i} lines is strongly affected by the residuals of the nebular subtraction. Therefore, the analysis of B268 focuses on lines lacking nebular components and the broader, lower order H\,{\sc i} lines, which show emission but since the width of the nebular emission is wider than the width between potential double peaks, we cannot distinguish the shape of the profile.

All H\,{\sc i} lines show photospheric absorption. For H$\beta$, H$\alpha$, Pa$\delta$, Pa$\gamma$, Pa$\beta$ and Br$\gamma$ this absorption is partially filled in by emission. For all of these lines, the blue side of the emission feature is stronger than the red side. The O\,{\sc i} line at 844.6\,nm is contaminated by nebular emission and the O\,{\sc i} triplet lines at 777.4\,nm and at 926.2-926.6\,nm are in absorption. [O\,{\sc i}] 630.0\,nm is single-peaked and in emission. 

The Ca\,{\sc ii} emission superimposed on the Paschen lines is strongly in emission. Only the blue peak of the feature can be used for analysis due to the blend. In contrast to B243, the Ca\,{\sc ii} triplet lines broader and stronger than the blended Paschen lines. The Fe lines are weak and in absorption, except for Fe\,{\sc ii} 999.8\,nm, which is single peaked and in emission. 
There is double-peaked Mg\,{\sc i} emission at 880.8\,nm and the Na\,{\sc i} doublet (D1 and D2) shows an additional absorption component on the red side beside the interstellar one. B268 shows the first and second overtone of the CO bandhead in emission. The bandhead profiles are characterized by a strong blue shoulder. 

\subsection{B275}
Photospheric absorption lines are observed in all H\,{\sc i} lines in the available wavelength range. Emission starts filling in this absorption from H$\delta$ to longer wavelengths and is clearly visible in both wings from H$\beta$. Similar to the previously discussed stars, the region around the rest wavelength of several lines is affected by residuals left from the subtraction of nebular emission. This region extends from 100-170\,km\,s$^{-1}$ (see B243 and Sec.~\ref{sec:neb} for comments on the width) for the Balmer series to 50-80\,km\,s$^{-1}$ for the Paschen series. While for some Paschen lines the double peaks are detected, for most lines it is impossible to distinguish between a double or a single peak. For all these H\,{\sc i} lines, the blue wing seems stronger than the red wing.

The stronger blue peak is also visible in the double-peaked O\,{\sc i} at 844.6\,nm. The O\,{\sc i} triplets at 777.3\,nm and 926.2-926.6\,nm appear in absorption, and there is single-peaked [O\,{\sc i}] emission at 630.0\,nm. The Ca\,{\sc ii} triplet appears double-peaked in emission and is stronger than the Paschen lines. While the blue feature shows two equally strong peaks, the redder ones have stronger red peaks. Many C\,{\sc i} and Fe lines show as double-peaked emission, varying in strength. Only Fe\,{\sc ii} 999.8\,nm exhibits a single peak. The Mg\,{\sc i} line at 880.8\,nm is observed in emission with a double-peaked profile. B275 shows first and second overtone CO bandhead emission. The profile shows a tiny blue shoulder. 

\subsection{B337}
Since B337 is the most embedded target, the spectrum is obscured for wavelengths shorter than 840\,nm, with an exception for the H$\alpha$ emission. For H$\alpha$, Pa$\delta$, Pa$\gamma$, Pa$\beta$ and Br$\gamma$ the emission is broader than the nebular emission, allowing for a disk detection but no further analysis. 

The spectra show a double-peaked Ca\,{\sc ii} triplet, double-peaked O\,{\sc i} emission at 844\,nm and Mg\,{\sc i} emission at 880.8\,nm, where the latter two lines only become apparent in the stacked spectrum. Fe\,{\sc ii} at 999.8\,nm is in emission. This peak is broad and because of the amount of noise in the spectrum it is not possible to distinguish between two peaks or one broad peak. The O\,{\sc i} triplets at 777.3\,nm and 926.2-926.6\,nm are in absorption and CO bandhead emission is weak, where the most clear detection is the second peak of the first overtone (at 2322.76\,nm). 

\newpage
\section{TVS of B215, B289 and B337}
         \begin{figure}[h]
   \resizebox{0.8\hsize}{!}
            {\includegraphics[width=0.9\textwidth]{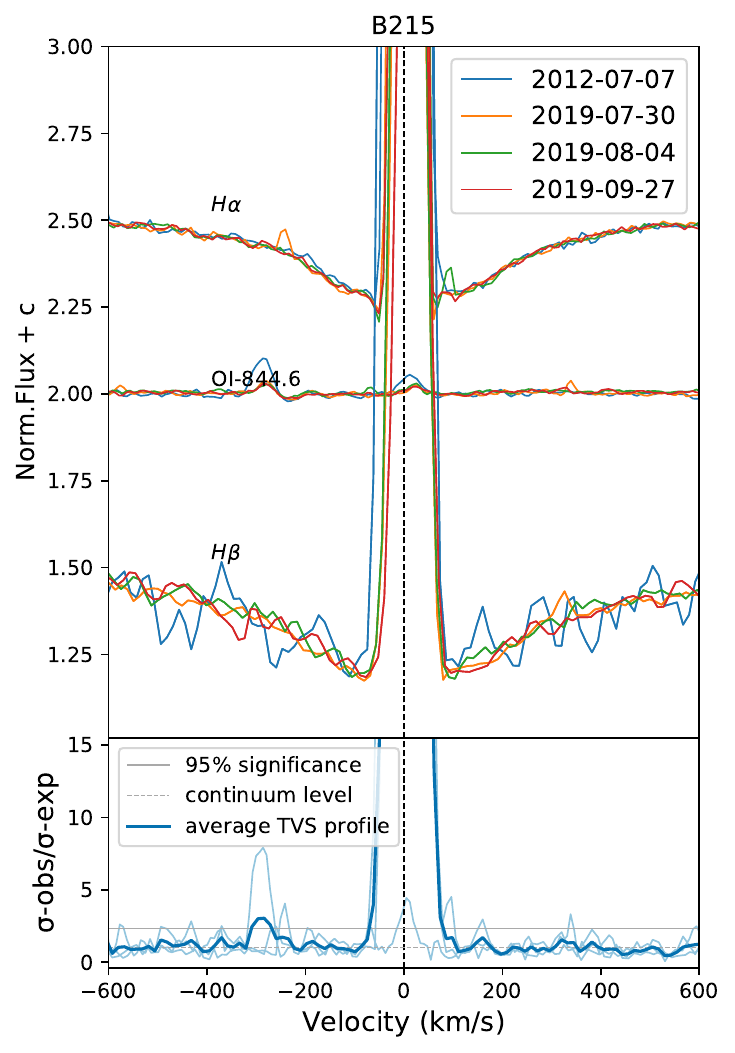}}
      \caption{Lines profiles and TVS of B215, see caption Fig.\,\ref{B243_TVS}. 
              }
         \label{B215_tvs}
   \end{figure}

            \begin{figure}[h]
   \resizebox{0.8\hsize}{!}
            {\includegraphics[width=0.9\textwidth]{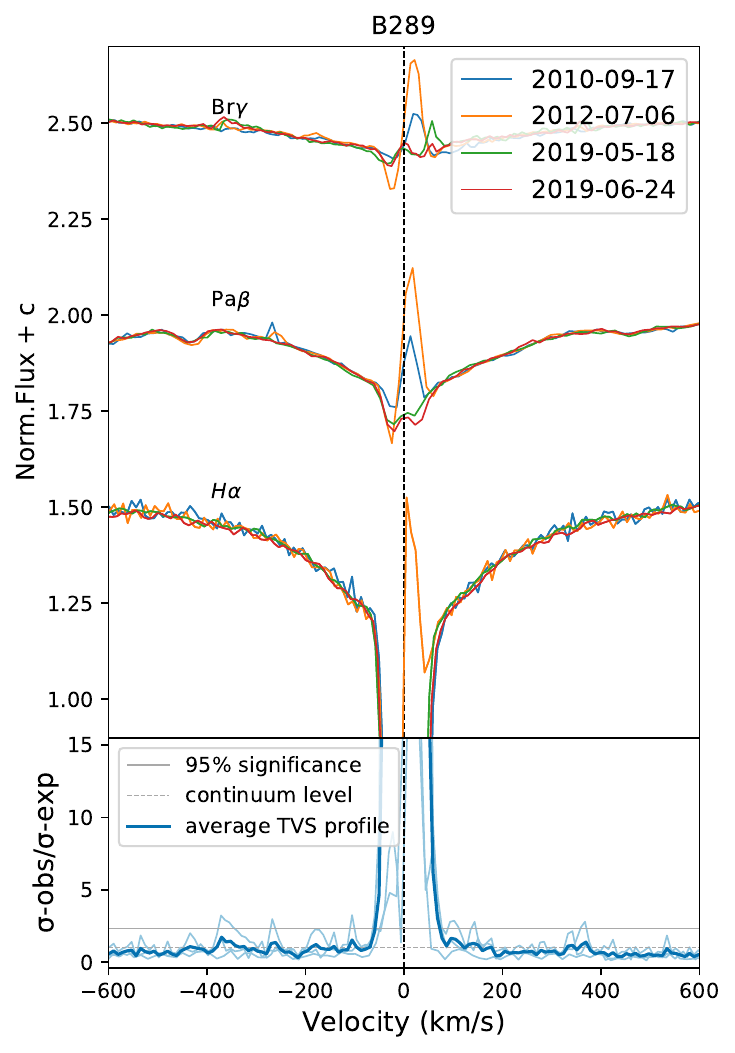}}
      \caption{Lines profiles and TVS of B289, see caption Fig.\,\ref{B243_TVS}.
              }
         \label{B289_tvs}
   \end{figure}
         \begin{figure}[h]
   \resizebox{0.8\hsize}{!}
            {\includegraphics[width=\textwidth]{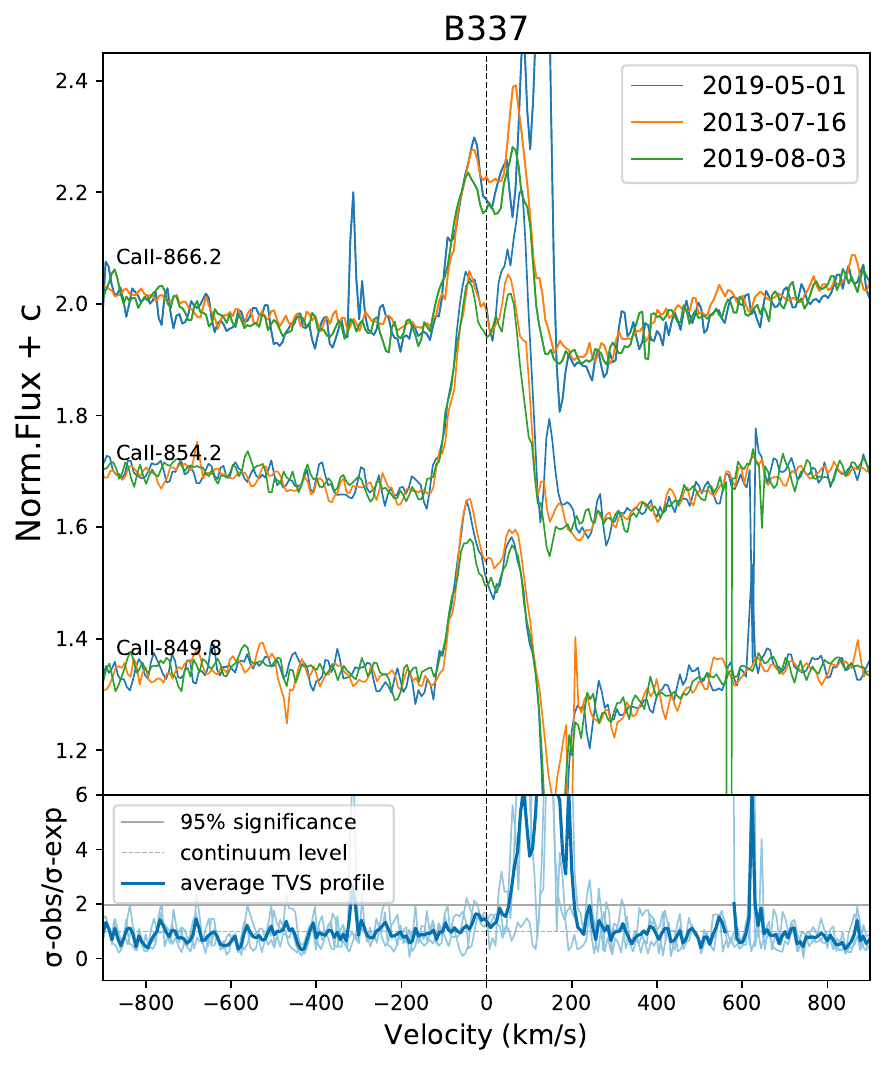}}
      \caption{Lines profiles and TVS of the Ca\,{\sc ii} lines in B337, see caption Fig.\,\ref{B243_TVS}.
              }
         \label{B337_tvs}
   \end{figure}
   
            \begin{figure}[h]
   \resizebox{0.8\hsize}{!}
            {\includegraphics[width=\textwidth]{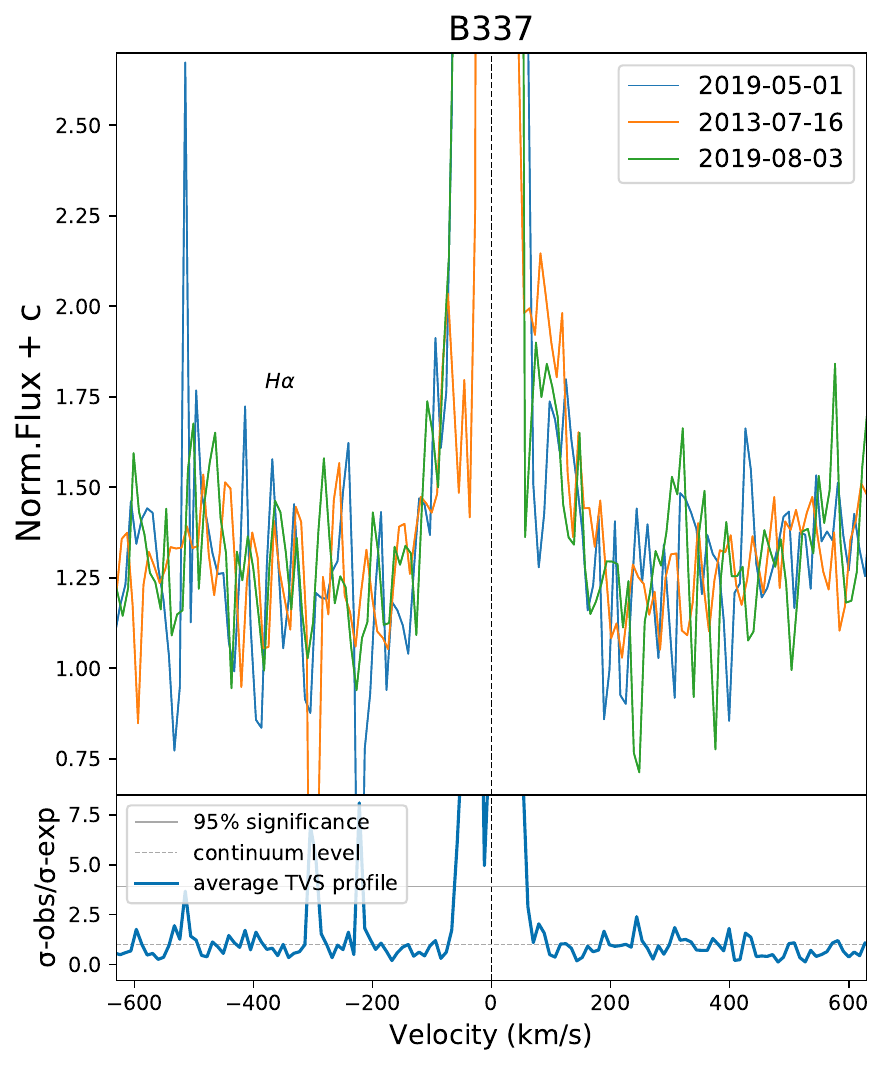}}
      \caption{Lines profiles and TVS of H$\alpha$ in B337, see caption Fig.\,\ref{B243_TVS}.
              }
         \label{B337_tvs_Halpha}
   \end{figure}

\newpage
\newpage

\section{EW and mass accretion rates}
In this section we present the EWs of a handful of lines and the mass accretion rates that follow from this that follow from the methods described in Sec.~\ref{sec:method_circumstellar} for the EW and in Sec.~\ref{sec:discus_massacc} for the mass accretion rates. In order to measure the EW of O\,{\sc i} at 844.6\,nm, which is blended with Pa-18, the average EW of Pa-17 and Pa-19 is subtracted from the EW of the blended O\,{\sc i} and Pa-18. The Ca\,{\sc ii} triplet is blended with Pa-16, Pa-15 and Pa-13 lines and the EW should therefore be taken as an upper limit.

\begin{sidewaystable}[p]
\caption{Mass accretion rates in M$_{\odot}$ year$^{-1}$}
\begin{tabular}{ll|llllll}
\midrule \midrule
     &                                & [O\,{\sc i}] & O\,{\sc i} & O\,{\sc i} & Ca\,{\sc ii} & CaII\,{\sc ii} & CaII\,{\sc ii} \\
     &                                & 630.0\,nm        & 777.4\,nm      & 844.6\,nm      & 849.8\,nm        & 854.2\,nm          & 866.2\,nm          \\
     \midrule \midrule
B243 & 2012-07-06 & 1.25\,x\,$10^{-6}$     & 1.46\,x\,$10^{-6}$   & 6.23\,x\,$10^{-6}$   &              &                &                \\
     & 2013-07-17 & 1.46\,x\,$10^{-6}$     & 2.26\,x\,$10^{-6}$   & 6.46\,x\,$10^{-6}$   &              &                &                \\
     & 2019-07-30 & 1.86\,x\,$10^{-6}$     & 5.30\,x\,$10^{-7}$   & 5.40\,x\,$10^{-6}$   &              &                &                \\
     & 2019-08-03 & 1.92\,x\,$10^{-6}$     & 2.83\,x\,$10^{-6}$   & 6.74\,x\,$10^{-6}$   &              &                &                \\
     & 2019-09-25 & 2.09\,x\,$10^{-6}$     & 3.04\,x\,$10^{-6}$   & 7.39\,x\,$10^{-6}$   &              &                &                \\
     \midrule
B268 & 2012-07-06 & 1.33\,x\,$10^{-6}$     &            &            & 5.83\,x\,$10^{-6}$     & 5.83\,x\,$10^{-6}$       & 3.44\,x\,$10^{-6}$       \\
     & 2013-07-17 & 8.60\,x\,$10^{-7}$      &            &            & 5.94\,x\,$10^{-6}$     & 5.94\,x\,$10^{-6}$       & 3.45\,x\,$10^{-6}$       \\
     & 2019-05-30                     & 1.71\,x\,$10^{-6}$     &            &            & 5.68\,x\,$10^{-6}$     & 5.68\,x\,$10^{-6}$       & 3.20\,x\,$10^{-6}$       \\
     & 2019-05-31                     & 1.58\,x\,$10^{-6}$     &            &            & 5.74x\,$10^{-6}$     & 5.74\,x\,$10^{-6}$       & 3.42\,x\,$10^{-6}$       \\
     & 2019-07-29                     & 1.74\,x\,$10^{-6}$     &            &            & 6.24\,x\,$10^{-6}$     & 6.24\,x\,$10^{-6}$       & 3.66\,x\,$10^{-6}$       \\
     \midrule
B275 & 2009-08-11                     & 1.04\,x\,$10^{-6}$     &            & 3.44\,x\,$10^{-6}$   & 6.31x\,$10^{-6}$     & 8.93x\,$10^{-6}$       & 4.76x\,$10^{-6}$       \\
     & 2019-06-05                     & 1.11\,x\,$10^{-6}$     &            & 4.24\,x\,$10^{-6}$   & 6.11\,x\,$10^{-6}$     & 8.49\,x\,$10^{-6}$       & 4.67\,x\,$10^{-6}$       \\
     & 2019-06-06                     & 1.05x\,$10^{-6}$     &            & 4.41x\,$10^{-6}$   & 6.19\,x\,$10^{-6}$     & 8.61\,x\,$10^{-6}$       & 4.65\,x\,$10^{-6}$       \\
     & 2019-07-09                     & 1.10\,x\,$10^{-6}$     &            & 4.05\,x\,$10^{-6}$   & 6.33\,x\,$10^{-6}$      & 8.71\,x\,$10^{-6}$       & 4.65\,x\,$10^{-6}$      \\
\bottomrule
\end{tabular}
\label{tab:mass_acc}
\end{sidewaystable}

\begin{sidewaystable}[p]
\caption{EW measurements of lines without nebular contamination in B243, B268 and B275 in \AA. In case of the Ca\,{\sc ii} the EWs are an upper limit}
\begin{tabular}{ll | lllllllll}
\midrule \midrule
 &
   &
  [O\,{\sc i}] &
  O\,{\sc i} &
  Pa-17 &
  Pa-18 &
  O\,{\sc i} &
  Pa-19 &
  Ca\,{\sc ii} &
  CaII\,{\sc ii} &
  CaII\,{\sc ii} \\
 &
   &
  630.0\,nm &
  777.4\,nm &
  846.7\,nm &
  843.8\,nm &
  844.6\,nm &
  841.3\,nm &
  849.8\,nm &
  854.2\,nm &
  866.2\,nm \\

\midrule \midrule
B243 &
  2012-07-06 &
-0.25$\pm$0.09	&
-0.38$\pm$0.10	&
-0.29$\pm$0.08	&
-0.27$\pm$0.10	&
-2.07$\pm$0.16	&
-0.24$\pm$0.06 &
   &
   &
   \\
 &
  2013-07-17 &
-0.29$\pm$0.14	& -0.6$\pm$0.12 &	-0.28$\pm$0.05	& -0.29$\pm$0.08 &	-2.15$\pm$0.18 &	-0.29$\pm$0.07 &
   &
   &
   \\
 &
  2019-07-30 &
-0.38$\pm$0.08	 & -0.13$\pm$0.16 &	-0.19$\pm$0.07	& -0.18$\pm$0.10 &	-1.77$\pm$0.17 &	-0.17$\pm$0.07 &
   &
   &
   \\
 &
  2019-08-03 &
-0.39$\pm$0.12  &	-0.76$\pm$0.13  &	-0.28$\pm$0.06  &	-0.25$\pm$0.1  &	-2.26$\pm$0.16  &	-0.22$\pm$0.07 &
   &
   &
   \\
 &
  2019-09-25 &
-0.43$\pm$0.08 &	-0.82$\pm$0.08 &	-0.44$\pm$0.07 &	-0.35$\pm$0.1 &	-2.5$\pm$0.17 &	-0.26$\pm$0.06 &
   &
   &
   \\
\midrule
B268 &
  2012-07-06 &
  -0.22$\pm$0.02 &
   &
   &
   &
   &
   &
-2.11$\pm$0.09 &	-2.65$\pm$0.06	& -2.65$\pm$0.09 \\
 &
  2013-07-17 &
  -0.14$\pm$0.02 &
   &
   &
   &
   &
   &
 -2.24$\pm$0.13 &	-2.7$\pm$0.08	 & -2.66$\pm$0.10 \\
 &
  2019-05-30 &
  -0.29$\pm$0.01 &
   &
   &
   &
   &
   &
  -1.96$\pm$0.09	& -2.59$\pm$0.08 &	-2.48$\pm$0.09 \\
 &
  2019-05-31 &
  -0.27$\pm$0.02 &
   &
   &
   &
   &
   &
-2.04$\pm$0.09	& -2.61$\pm$0.07	 & -2.64$\pm$0.09 \\
 &
  2019-07-29 &
  -0.30 $\pm$0.02 &
   &
   &
   &
   &
   &
 -2.2$\pm$0.11 &	-2.83$\pm$0.10 &	-2.81$\pm$0.13 \\
\midrule
B275 &
  2009-08-11 &
-0.17$\pm$0.03	& &	-1.03$\pm$0.06	& -0.71$\pm$0.10 &	-0.90$\pm$0.25	& -0.38$\pm$0.05 &	-2.43$\pm$0.04	& -4.01$\pm$0.07 &	-3.58$\pm$0.06 \\
 &
  2019-06-05 &
  -0.18$\pm$0.04	& &	-0.98$\pm$0.07 &	-0.71$\pm$0.08 &	-1.13$\pm$0.18 &	-0.43$\pm$0.03 &	-2.34$\pm$0.04 &	-3.82$\pm$0.05 &	-3.53$\pm$0.03 \\
 &
  2019-06-06 &
  -0.17$\pm$0.03	& &	-1.02$\pm$0.08 &	-0.73$\pm$0.10 &	-1.19$\pm$0.22 &	-0.43$\pm$0.05 &	-2.38$\pm$0.04 &	-3.87$\pm$0.06 &	-3.51$\pm$0.04 \\
 &
  2019-07-09 &
 -0.18$\pm$0.04	& &	-0.98$\pm$0.07 &	-0.70$\pm$0.09 &	-1.08$\pm$0.22 &	-0.42$\pm$0.05 &	-2.43$\pm$0.05 &	-3.91$\pm$0.06 &	-3.51$\pm$0.07 \\
\bottomrule
\end{tabular}
\label{tab:EW}
\end{sidewaystable}

\section{Peak-to-peak, V/R-ratio and (V/R)$_{wing}$-ratio} [p]
This section shows the peak-to-peak velocities and (V/R)$_{wing}$-ratio for each star in each epoch. Relative changes over time are discussed in Sec.~\ref{p1:sec:DP_emission} and \ref{sec:varia_wing}. The V/R-ratio for each epoch in B243 are shown for comparison with changes in (V/R)$_{wing}$-ratio, that are largest for this star. 

         \begin{figure}[p]
   \resizebox{0.75\hsize}{!}
        {\includegraphics[width=0.75\textwidth,clip]{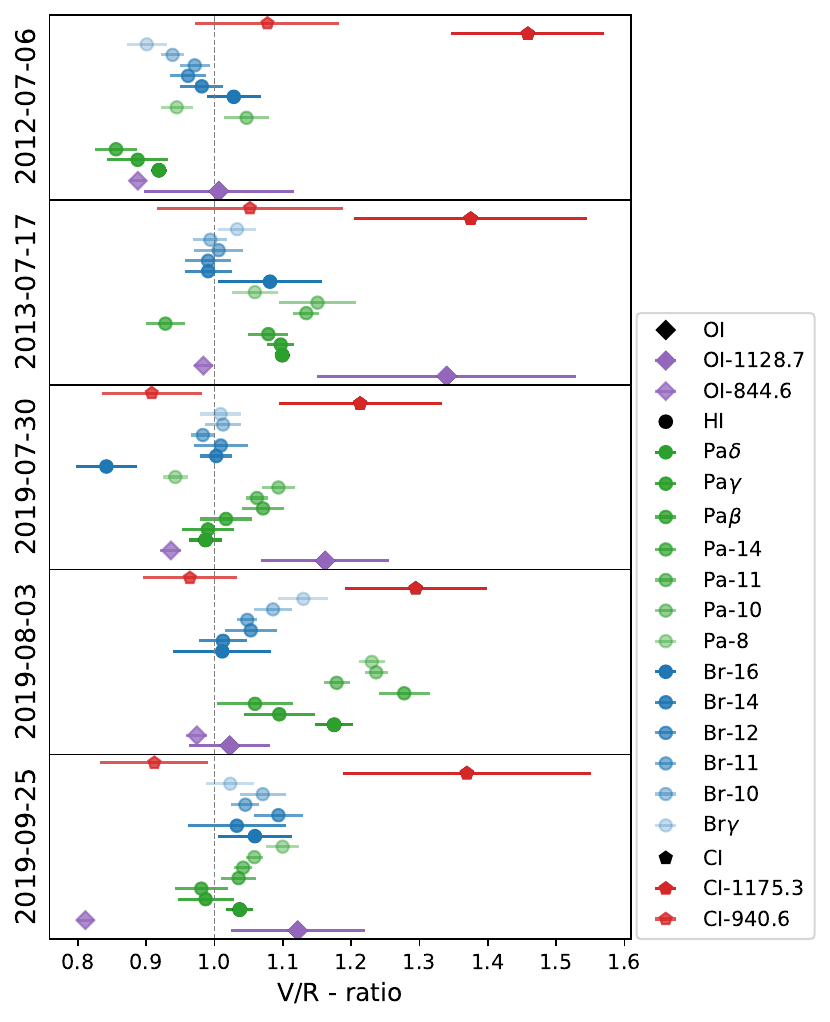}}
      \caption{V/R-ratio between the peaks of the double-peaked emission features in B243. Most lines follow a similar trend in ratio over the several epochs. Part of the lines are <\,1 in the first three epochs, indicating a stronger red peak than blue peak. This changes for the last two epochs where most of the lines have a stronger blue peak than red peak. 
              }
         \label{B243_VR_peak}
   \end{figure}
   
    \begin{figure*}[p]
   \centering
   \includegraphics[width=\hsize]{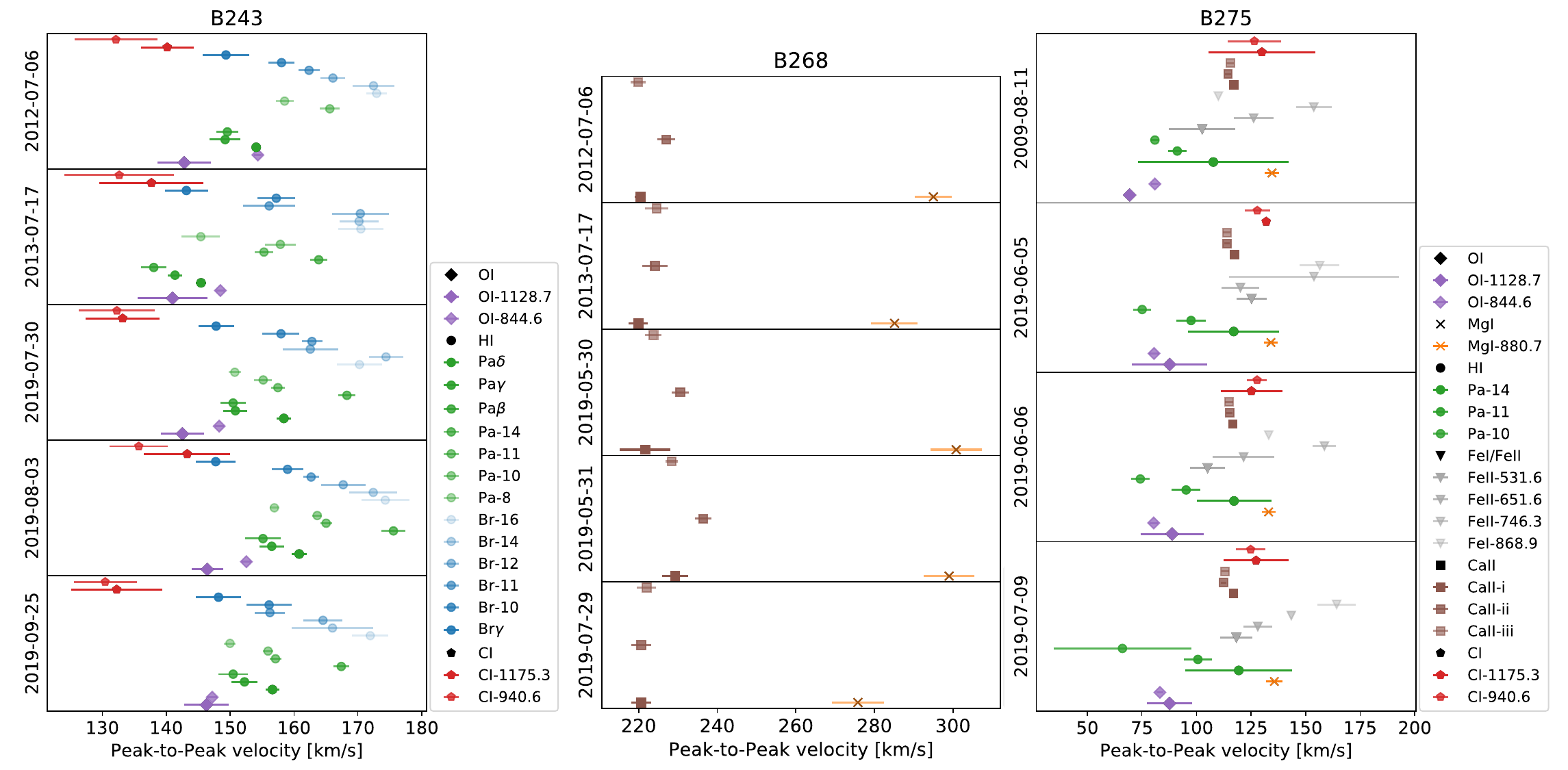}
      \caption{The peak-to-peak velocity of the double-peaked emission lines in the spectra of B243, B268 and B275 shows on the x-axis and the date of observation on the y-axis. The peaks are measured after subtracting the stellar feature, by double Gaussian fit. Velocities in B243 (left figure) range between 130\,km\,s$^{-1}$ and 180\,km\,s$^{-1}$ and there is a general trend in velocity over the five epochs, which is most apparent for all lines in the last three epochs. In B268 (middle figure) the velocities range between 220\,km\,s$^{-1}$ and 300\,km\,s$^{-1}$, where there is a velocity difference of ~65\,km\,s$^{-1}$ between the Ca\,{\sc ii} triplet and Mg\,{\sc i} peak-to-peak velocities (the legend of B275 can also be used to read the middle figure). Velocities in B275 (right figure) range from 70\,km\,s$^{-1}$ and 165\,km\,s$^{-1}$.}
         \label{fig:peaktopeak}
   \end{figure*} 
   \begin{figure*}[h]
   \centering
   \resizebox{\hsize}{!}
            {\includegraphics[width=\textwidth,clip]{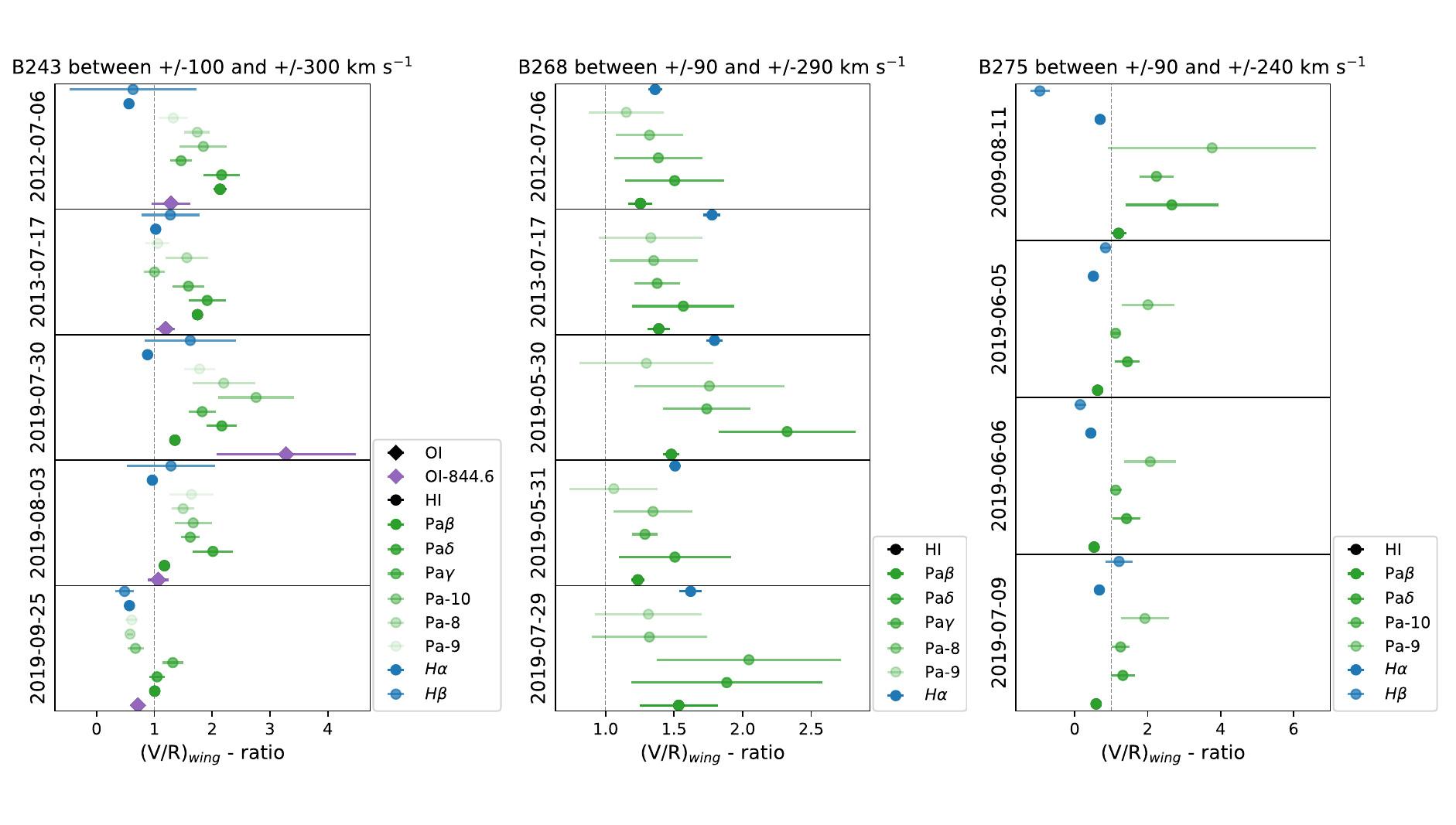}}
      \caption{(V/R)$_{wing}$-ratio measurements of the double-peaked emission features after subtracting a star model is on the x-axis and the observation date on the y-axis. In the left panel for B243, the EW is determined between $\pm$100\,km\,s$^{-1}$ and $\pm$300\,km\,s$^{-1}$; in the middle, for B268, between $\pm$90\,km\,s$^{-1}$ and $\pm$290\,km\,s$^{-1}$, and on the right, for B275, between $\pm$90\,km\,s$^{-1}$ and $\pm$240\,km\,s$^{-1}$.
              }
         \label{B243_VR}
         \label{B268_VR}
         \label{B275_VR}
   \end{figure*}

\end{appendix}
\end{document}